\documentclass[twocolumn,prb,multicol,amsmath,amssymb]{revtex4}
\usepackage[dvips]{graphicx}
\usepackage{graphicx}
\usepackage{dcolumn}
\usepackage{bm}
\usepackage{graphics}
\usepackage{epsfig,color}
\usepackage[normalem]{ulem} 
\usepackage{soul}

\newcommand{\be}{\begin{equation}}
\newcommand{\ee}{\end{equation}}
\newcommand{\bea}{\begin{eqnarray}}
\newcommand{\eea}{\end{eqnarray}}

\begin{document}
\title {Achieving spin-squeezed states by quench dynamics in a quantum chain}
\author{Hadi Cheraghi$^{1,2,3}$}
\email[]{h.cheraghi1986@gmail.com}
\author{Saeed Mahdavifar$^{1}$}
\author{Henrik Johannesson$^{2}$}
\affiliation{$^{1}$Department of Physics, University of Guilan, 41335-1914, Rasht, Iran}
\affiliation{$^{2}$Department of Physics, University of Gothenburg, SE 412 96 Gothenburg, Sweden}
\affiliation{ $^{3}$Institute of Physics, Maria Curie-Sk\l odowska University, Plac Marii Sk\l odowskiej-Curie 1, PL-20031 Lublin, Poland}  
\begin{abstract}
We study the time evolution of spin squeezing in the one-dimensional spin-1/2 XY model subject to a sudden quantum quench of a transverse magnetic field. The initial state is selected from the ground state phase diagram of the model, consisting of ferro- and paramagnetic phases separated by a critical value of the transverse field. Our analysis, based on exact results for the model, reveals that by a proper choice of protocol, a quantum quench from an unsqueezed state can create spin squeezed nonequilibrium states. We also identify a nonanalyticity in the long-time average of the spin-squeezing parameter when quenching to the equilibrium quantum critical point. This suggests that the ferro- and paramagnetic phases also define distinct phases for how the transverse field redistributes  quantum fluctuations among the spin components away from equilibrium.
\end{abstract}
\maketitle

\section{Introduction}

As is well known, non-commuting observables in quantum systems are subject to intrinsic uncertainty.
The Heisenberg uncertainty principle puts a limit on the product of variances of such observables $-$ the different components of a spin operator being prime examples. A minimum-uncertainty state where all components of the {\color{black} collective spin operator of an ensemble of spinful particles} have the same variance is called a coherent spin state [\onlinecite{Radcliffe}]. The phenomenon of spin squeezing refers to the redistribution of quantum fluctuations in a minimum-uncertainty state whereby the variance of one of the spin components becomes smaller than those of the other components [\onlinecite{Kitagawa,Wineland,Ma1}]. 

The study and experimental use of spin squeezing have attracted a lot of interest in recent years. Examples of the latter include spin squeezing as a tool to produce [\onlinecite{Sorensen}] and detect [\onlinecite{Wang,Toth,Guehne}] quantum entangled states.  Within the field of quantum metrology [\onlinecite{Giovannetti}], spin squeezing has figured prominently as a means to improve on atomic precision measurements beyond the standard quantum limit (defined as the maximum phase sensitivity achievable with separable states) [\onlinecite{Wineland,Hald,Orzel,Meyer,Smith,Takano,Gross1,Riedel,Hamley,Auccaise,Hosten,Orioli,Pezze,Braverman,Bao}].
The last two decades have also seen advances in the theoretical understanding of various aspects of spin squeezing [\onlinecite{Sorensen1,Duan,SorensenNature,Law,Wang2,Sorensen2,Wang1,Ma2,Wang3,Liu,Abad,Xu1,Frerot,Balazadeh,Kaubruegger,Schulte,Qin,Comparin2}] $-$ from its creation by interactions in optical lattices [\onlinecite{Sorensen1}] to its use in protocols for programmable quantum sensors [\onlinecite{Kaubruegger}]. Insights thus gained touch on fundamentals, such as the recognition that extreme spin squeezing appears at the quantum critical point of the Ising model in a transverse field, signaling an enhanced growth of entanglement at quantum criticality [\onlinecite{Frerot}].

{\color{black} The prototype setting for spin squeezing is modeled by the one-axis twisting (OAT) Hamiltonian [\onlinecite{Kitagawa,Ma1}], built from uniform, infinite-range Ising interactions. Experimental implementations include collisional interactions between delocalized atoms [\onlinecite{Zibold,Martin}], and interactions mediated by coupling to phonons [\onlinecite{Britton,Bohnet}] or cavity modes [\onlinecite{Baumann,Ritsch,Davis}]. In an idealized protocol, the system is initialized in a coherent state of $N$ polarized spins. The OAT Hamiltonian is then applied, causing a shear of the Gaussian-type spin distribution, leading to a spin-squeezed state with a reduced variance along some axis. The dynamics controlled by the OAT Hamiltonian can be solved analytically, making this model widely used and studied. However, experimental platforms relying on uniform infinite-range interactions come with their own challenges, spurring interest in more easily controllable realizations {\color{black} or simulations} of OAT-type models where the spin interactions fall off with distance as a power-law [\onlinecite{Hazzard,Feig,Lerose}] {\color{black} or extend only to nearest neighbors [\onlinecite{Gietka}].} Of special interest is to find out  under what conditions short-range {\color{black} or power-law} interactions can produce spin squeezing that scales with the system size [\onlinecite{Perlin,Comparin1}], with an eye towards applications with cold atomic gases [\onlinecite{Cazalilla,Gross2}], Rydberg atoms [\onlinecite{Browaeys}], and trapped ions [\onlinecite{Britton}]. This line of research also aims to find out about the dynamics of spin squeezing, largely motivated by advances in experiments on cold atoms and trapped ions out of equilibrium [\onlinecite{Eisert}].}
 
Here we try to contribute to this effort by studying the time evolution of spin squeezing realized in a spin-1/2 XY chain in a transverse magnetic field after a sudden quench, {\color{black} with the {\color{black} transverse field} Ising chain contained in the model as a special case. Different from the standard spin-squeezing protocol where the system is initialized in a coherent state [\onlinecite{Ma1}], we conceive the system to be prepared in the ground state of the initial Hamiltonian $-$ which may or may not be coherent. After a sudden change of the strength of the magnetic field,} the presence of spin squeezing is looked for as the state evolves with time. Our choice of the XY chain as a case study is motivated by its simplicity, allowing for a well-controlled study, but also because the model is integrable. This means that it does not thermalize once having been taken out of equilibrium [\onlinecite{Barouch1,Igloi}]. This is so, since the quasiparticles excited by a quench do not follow the standard Gibbs distribution, but rather a generalized Gibbs ensemble (GGE) [\onlinecite{Essler}]. While not a main theme of our work, it does provide a link to future studies of spin-squeezing dynamics in the realm of  GGE physics.

The equilibrium phase diagram of the XY chain at zero temperature exhibits two regions: a ferromagnetic phase with a spontaneously broken $Z_2$ symmetry in the thermodynamic limit, separated from a {\color{black} paramagnetic phase} by a quantum phase transition at a critical value $h=h_c$ of the magnetic field [\onlinecite{Katsura,Niemeijer}]. Depending on the strength of the magnetic field $h$ and the amount of anisotropy of the spin interaction, measured by a parameter $\delta$, the ground state can further be classified as being unsqueezed, spin coherent or spin squeezed. By studying the dynamics of spin squeezing we show that by properly adjusting the control parameter $h$ (keeping $\delta$ fixed) we are able to {\color{black} achieve squeezed} states from a quantum quench even when the initial state is neither coherent nor squeezed.

Our results reveal that the {\color{black} long-time average $\overline{\xi _s^{2}}$} of the parameter which quantifies the amount of time-dependent spin squeezing in the system $-$ the {\em spin-squeezing parameter} {\color{black} $\xi_s^2(t)$} [\onlinecite{Kitagawa}] $-$ {\color{black} exhibits a nonanalyticity at the equilibrium} quantum critical point $h\!=\!h_c$. This result $-$ {\color{black} which suggests that the way the transverse field redistributes quantum fluctuations among the spin components undergoes a phase transition at $h\!=\!h_c$} $-$ ties in to a recent finding of a nonanalyticity in the spin-squeezing parameter at the quantum critical point of the Lipkin-Meshkov-Glick model [\onlinecite{Xu2}].   

We also establish that there is a universality in the revival times for the spin squeezing in a finite system. Specifically, the revival times do not depend on the initial state or the size of the quench and are given by integer multiples of the revival period $T_{rev} \simeq {N}/{2v_{max }}$, with $N$ the system size and $v_{max }$ the maximal group velocity of quasiparticles excited by the quench. This mirrors known results for revival times for  the Loschmidt echo [\onlinecite{Hamma,Henrik,Delgado}] and maximal quantum Fisher information [\onlinecite{Mishra,Hadi2,Akbari}] for quenched quantum many-body systems.

The rest of the paper is laid out as follows: In Sec.~II we present the model and review its exact solution. Section~III presents the key formulas and theoretical approach that underpin our analysis. Results, with numerical plots illustrating the {\color{black} dynamics of the spin-squeezing parameter for} various quench scenarios, are {\color{black} presented in} Sec.~IV. {\color{black} This section, which is separated into three parts, corresponding to squeezed, unsqueezed and spin-coherent initial states, also contains numerical results for the time-dependence of the variance of the mean spin direction.} Section V presents results for the revival structure of the spin-squeezing parameter, with focus on the universality of the revival time. Section VI, finally, contains a summary and outlook.

\section{XY chain in a transverse field}
The Hamiltonian of the one-dimensional spin-1/2 XY model in the presence of a transverse magnetic field is given by   
\begin{eqnarray}\label{eq1}
{\cal H} =&-&J\sum\limits_{n = 1}^N {\left( {(1\! + \!\delta )S _n^x S_{n \!+ \!1}^x + (1 \!- \!\delta ) S_n^y S_{n \!+\! 1}^y} \right)} \nonumber \\
&- &h\sum\limits_{n \!= \!1}^N {S_n^z},
\end{eqnarray}
where the components of the spin operators $S_n^{\mu}$ on site $n$ are represented by Pauli matrices, $S_n^\mu = \sigma_n^\mu/2$ for $\mu = x, y, z$, $J>0$ denotes the ferromagnetic exchange coupling,  and $\delta$ and $h$ are the anisotropy parameter and magnitude of the magnetic field, respectively. We here consider the case of periodic boundary conditions, $S_{n+N}^\mu=S_n^\mu$, with $N$ the number of lattice sites. We further restrict the anisotropy parameter to the interval $(0,1]$, with $\delta=1$ the Ising chain in a transverse field. Note that we have excluded the case $\delta=0$, corresponding to the gapless XX model, not to be addressed in this work.

The model exhibits a quantum phase transition in the thermodynamic limit $N \rightarrow \infty$ at $h_c \!= \!J$, from a ferromagnetic (FM) phase ($h \!< \!J$) to a paramagnetic (PM) spin-polarized phase ($h\!>\!J$) [\onlinecite{Katsura}]. When the system is finite, the FM phase is characterized by intermediate-range (long-range) longitudinal spin correlations for $0 \!< \!h \!< \!J$ ($h\!=\!0$), while "true" macroscopic ordering, with a finite value of the longitudinal magnetization, sets in only in the thermodynamic limit where the $Z_2$ symmetry (invariance under $\pi$-rotations around the $z$-axis) gets spontaneously broken. Intriguingly, the finite-size exponentially small splitting of the two-fold degenerate ground state at $h=0$, caused by tunneling driven by the transverse field when $h\neq 0$, collapses on the circle $h^2+(J \delta) ^2=J^2$ where the degeneracy is fully restored [\onlinecite{Kurmann}]. Here the ground state gets factorized into a direct product of single spin states [\onlinecite{Adesso}] $-$ also known as a coherent spin state [\onlinecite{Radcliffe}]$-$ with implications for the spin squeezing, to be discussed below. 

The Hamiltonian in (\ref{eq1}) is integrable and, as first shown by Lieb {\em et al.} [\onlinecite{LSM}], in the absence of a magnetic field can be mapped onto a system of free spinless fermions by a Jordan-Wigner (JW) transformation,
\begin{eqnarray} \label{JWT}
\sigma_n^{+} &=& \prod_{\l=1}^{n-1}(1-2a^\dagger_{\l} a_{\l})a_n, \ \ \  \sigma_n^{-} = \prod_{\l=1}^{n-1}(1-2a^\dagger_{\l} a_{\l})a_n^\dagger, \nonumber \\
 \sigma_n^z &=&  2a_n^\dagger a_n - 1, 
\end{eqnarray}
where, as usual, $\sigma^{\pm}_n = (\sigma^x_n \pm i\sigma^y_n)/2$, and $a_n^\dagger$ and $a_n$ are the fermionic operators. One thus obtains 
\begin{eqnarray}\label{eq2}
{\cal H} &=& -\frac{J}{2}  \sum\limits_{n = 1}^N ( a_n^\dag a_{n + 1}+\delta a_n^\dag a_{n + 1}^\dag  + \mbox{H.c.} ) \nonumber\\
& -& h\sum\limits_{n = 1}^N {\left( {a_n^\dag {a_n} - 1/2} \right)}.
\end{eqnarray}
We have here discarded the boundary term induced by the JW transformation since it contributes only to ${\cal O}(1/N)$ in the energy spectrum and hence is negligible for large $N$ [\onlinecite{FranchiniBook}]. By this, we study what is known in the literature as the {\em c-cycle} problem of the XY model [\onlinecite{LSM}]. Performing a Fourier transformation ${a_n} =  \sum_k e^{ - ikn} {a_k}$, followed by a Bogoliubov transformation 
${a_k} = \cos ({\theta _k}) {\alpha _k} + i\sin ({\theta _k}) \alpha _{ - k}^{\dag}$, yields the quasiparticle Hamiltonian  
\begin{equation}\label{eq3}
{\cal H} = \sum\limits_k {{\varepsilon }_k( {\alpha _k^\dag {\alpha _k} - 1/2} )},
\end{equation}
with energy spectrum 
\begin{equation} \label{spectrum}
\varepsilon_k = \sqrt{{\cal A}_k ^2+{\cal B}_k ^2},
\end{equation}
where
${\cal A}_k= [\cos(k)+h]$ and ${\cal B}_k=  \delta \sin(k)$
are related to the Bogoliubov angle $\theta_k$ by $\tan (2{\theta _k}) = {\cal B}_k/{\cal A}_k$. The summation in Eq. (\ref{eq3}) runs over $k=2\pi m/N$, with $m=0,\pm 1,...,\pm \frac{1}{2}(N-1) \ [m= 0, \pm 1,..., \pm (\frac{1}{2}N-1), \frac{1}{2}N]$ for $N$ odd [$N$ even] (having imposed periodic boundary conditions on the JW fermions). Here, and in what follows, we have put $J=1$.

As we shall see in the next section, all information about the dynamics of the spin squeezing can be extracted from the spectrum in Eq. (\ref{spectrum}), using known formulas for the spin correlation functions of the model.    \\ 

\section{Dynamical Spin Squeezing}
\subsection{Spin-squeezing parameter}
A simple way to quantify the amount of spin squeezing in a state {\color{black} $|\psi(t) \rangle = e^{-i{\cal H}t}|\psi(0) \rangle$ time-evolved by the Hamiltonian ${\cal H}$} in (\ref{eq1}) is suggested by the spin-uncertainty relation [\onlinecite{Ma1}], 
\begin{equation} \label{spinuncertainty}
(\Delta J_{\alpha})^2 (\Delta J_{\beta})^2 \ge |\langle J_{\gamma} \rangle|^2/4,
\end{equation}
with $\Delta J_{\alpha} = \sqrt{\langle J_{\alpha}^2\rangle - \langle J_{\alpha} \rangle^2}$, and with the expectation values calculated for 
\begin{equation} \label{Jdef}
J_{\alpha} \!= \!\sum_{n=1}^{N} {S _n^ \alpha}, \ \ \alpha\!=\!x,y,z,
\end{equation}
and similarly for $\Delta J_{\beta}$ and $\Delta J_{\gamma}$. {\color{black} From here on, and for brevity, we write time-dependent expectation values and equal-time spin correlation functions using the notation $\langle \ldots \rangle := \langle \psi(t) | \ldots | \psi(t) \rangle$ (unless otherwise stated).} 

{\color{black} From the above one} may conclude that squeezing is present as soon as one of the fluctuations on the left-hand side of (\ref{spinuncertainty}) satisfies $(\Delta J_{\alpha})^2 < |\langle J_{\gamma}\rangle|/2$, implying a squeezing parameter $\xi_s^2 = 2(\Delta J_{\alpha})^2/|\langle J_{\gamma}\rangle | \ ( \alpha \neq \gamma$) which signals a squeezed state when $\xi_s^2 < 1$. However, as is easily verified, a change of basis may spuriously lead to $\xi_s^2 < 1$ also for a coherent spin state. To amend for this, Kitagawa and Ueda [\onlinecite{Kitagawa}] defined an improved squeezing parameter $\xi_s^2$,
with a ``built-in" fixed reference direction, reflecting the fact that fluctuations in a coherent spin state are to be defined with respect to the mean spin direction 
$\vec{n}_0 = \langle \vec{J} \rangle/|\langle \vec{J} \rangle|$:
\begin{eqnarray} \label{eq5}
\xi _s^2 = \frac{{4{{({\Delta {J_{\vec{n}_{\perp}}}})}^2}}}{N}.
\end{eqnarray}   
Here the subscript $\vec{n}_{\perp}$ defines the direction perpendicular to $\vec{n}_0$ where the minimal value of the
variance $(\Delta J)^2$ is obtained, with $J_{\vec{n}_{\perp}}\! = \!\vec{J}\cdot \vec{n}_{\perp}$. A coherent spin state corresponds to $\xi _s^2 =1$, with the inequality $\xi _s^2 <1$ indicating that the system is spin squeezed. Let us mention that the definition of spin-squeezing parameter in (\ref{eq5}) is not unique; other definitions have been proposed, e.g., in Ref. [\onlinecite{Wineland}], adapted to Ramsey spectroscopy experiments. In this work we apply the construction in (\ref{eq5}), usually favored in theoretical studies of model systems.

Let us now go back to the XY model in (\ref{eq1}) and extract an expression for $\xi _s^2$ to be used in our calculations. The unbroken $Z_2$ invariance for finite $N$ (to be assumed from now on) implies that the in-plane magnetization vanishes,
\begin{eqnarray}\label{eq6}
\left\langle {J_x } \right\rangle  = \left\langle {J_y } \right\rangle  = 0,
\end{eqnarray}
and similarly,
\begin{equation} \label{Jdef}
\left\langle {J_\alpha J_z} \right\rangle  = \left\langle {J_z J_\alpha } \right\rangle  = 0, \ \ \alpha  = x,y.
\end{equation}
It follows that the magnetization for $h>0$ is always along the $z$-direction, with full polarization developing {\color{black} for large $h$} in the PM phase $h > h_c$. Consequently, 
$J_{\vec{n}_{\perp}} = \cos (\Omega )J_x + \sin (\Omega)J_y$, with $\Omega$ to be chosen so as to minimize
\begin{eqnarray}\label{eq7}
({\Delta {J_{\vec{n}_{\perp}}}})^2 &=& \left\langle (J_ {\overrightarrow n  \bot } )^2 \right\rangle  - \left\langle J_ {\overrightarrow n  \bot }  \right\rangle ^2 \nonumber \\
&=& \langle (\cos(\Omega)J_x + \sin(\Omega)J_y)^2\rangle,
\end{eqnarray}
where the second identity follows from Eq. (\ref{eq6}). {\color{black} Combining Eqs. (\ref{eq5}) and (\ref{eq7}) one finds that
\begin{eqnarray} \label{eq8}
\xi _s^2 &=& \frac{2}{N}\mathop {\min }\limits_\Omega  \Big( {\left\langle {J_x^2 + J_y^2} \right\rangle } + \cos (2\Omega )\left\langle {J_x^2 - J_y^2} \right\rangle \nonumber \\ 
& & \ \ \ \ \ \ \ \ \ \ \ \ + \sin(2\Omega)\langle J_xJ_y + J_yJ_x\rangle \Big) \nonumber \\
& = & \frac{2}{N}\Big( {\left\langle {J_x^2 \!+\! J_y^2} \right\rangle  \!-\! \sqrt {{{\left\langle {J_x^2 \!-\! J_y^2} \right\rangle }^2} \!+\! {{\left\langle {{J_x}{J_y} \!+\! {J_y}{J_x}} \right\rangle }^2}} } \Big).
\nonumber \\   
\end{eqnarray}
Given Eq. (\ref{Jdef}),
\begin{equation} \label{spinspin}
\langle {J_\alpha }{J_\beta}\rangle = N\langle {S_1^\alpha S_1^\beta }\rangle  + N\sum\limits_{n = 1}^{N - 1} G_n^{\alpha \beta }, \ \ \ \alpha, \beta = x, y, z
\end{equation}
with $G_n^{\alpha \beta } := \langle S_1^{\alpha} S_{1+n}^{\beta}\rangle$ an equal-time spin correlation function and where we have taken advantage of translational invariance to simplify the expression. Using that
$\langle S_1^{\alpha} S_1^{\alpha}\rangle = 1/4$, Eqs. (\ref{eq8}) and (\ref{spinspin}) yield a closed expression for the spin squeezing parameter $\xi_s^2(t)$ in terms of the correlation functions $G_n^{\alpha \beta}(t)$:
\begin{multline} \label{xiclosed}
\xi _s^2(t) =  1+ 2\sum\limits_{n = 1}^{N - 1} ( G_n^{xx}(t) + G_n^{yy}(t)) \\ 
- \! 2\sqrt { \!\big[\sum\limits_{n = 1}^{N - 1} \!(G_n^{xx}(t) \!-\! G_n^{yy}(t)) \big] ^2 \!+\!\big[\sum\limits_{n = 1}^{N - 1} \!( G_n^{xy}(t) \!+\! G_n^{yx}(t)) \big]^2\! }  
\end{multline}}
To {\color{black} remind the reader that we are} exploring the out-of-equilibrium dynamics after a quantum quench, we have expressly inserted the time argument in Eq. (\ref{xiclosed}). 

\subsection{Quench protocols}
Quantum simulators have provided experimental access to the real-time dynamics of quantum matter at an unprecedented
level of control and now make possible high-precision studies of the nonequilibrium dynamics of isolated quantum systems [\onlinecite{Eisert,Schreiber,Choi,Gorg}]. Here quantum quenches play an important role [\onlinecite{Mitra}]. 
In the standard quench protocol, the system is prepared in an eigenstate $\left| {{\Psi _0}} \right\rangle $ (usually the ground state) of some Hamiltonian, call it $\mathcal{H}(h_1)$, with $h_1$ the value of a tunable control parameter. The system is then taken out of equilibrium by a sudden change of the control parameter from its initial value $h_1$ to a different value $h_2$, yielding a post-quench Hamiltonian ${\cal H}(h_2)$ which now governs the time evolution, $\left| {\Psi (t)} \right\rangle  = {e^{ - i\mathcal{H}(h_2)t}}\left| {{\Psi _0}} \right\rangle$. 
Here we shall do quantum quenches on the Hamiltonian ${\mathcal H(h)}$ in (\ref{eq1}), using the magnetic field $h$ as control parameter and choosing different initial states, all being ground states of ${\mathcal H(h)}$. Depending on the particular choice of $h=h_1$, an initial state may be coherent, squeezed, or unsqueezed, as diagnosed by the value of the spin-squeezing parameter in (\ref{xiclosed}) {\color{black} when $t\!=\!0$.} The resulting spin-squeezing dynamics after a quench {\color{black} $h_1 \rightarrow h_2$, coded by} $\xi_s^2(t)$, is then monitored by numerically {\color{black} plotting} the time-dependence of the correlation functions in (\ref{xiclosed}). For this we need some more results, to be reviewed next.
\subsection{Dynamics of spin squeezing}
To determine $\xi_s^2(t)$ in eq. (\ref{xiclosed}), we need to calculate the two-point functions {\color{black} $G ^{\alpha \beta}_n(t)  : = \langle {S _1^\alpha (t) S _{1+n}^\beta (t)} \rangle$, with 
$\alpha, \beta = x, y$.} 
Because of the nonlocal nature of the Jordan-Wigner transformation, this calculation {\color{black} is nontrivial. For the diagonal two-point functions $G_n^{xx}(t)$ and $G_n^{yy}(t)$ one can rely on known results for the time-independent case where the calculation is reduced to one of Toeplitz determinants [\onlinecite{LSM,Barouch2}].  By time evolving the fermionic correlators that make up these determinants, one obtains exact expressions for $G_n^{xx}(t)$ and $G_n^{yy}(t)$. The calculation of the off-diagonal two-point functions $G_n^{xy}(t)$ and $G_n^{yx}(t)$ can be carried out in a similar vein, also here using the Wick theorem to express correlators of fermion operators as Pfaffians [\onlinecite{Fubini}]; for details see the Appendix.

To illustrate the core of the rather long analysis in the Appendix, let us consider one of the off-diagonal two-point functions, say $G_n^{xy}(t)$. Introducing $A_n = a_n^{\dagger}+a_n$ and $B_n = a_n^{\dagger}-a_n$ with $a_n^{\dagger}$ and $a_n$ the JW fermions in Eq. (\ref{JWT}), a direct calculation yields an expression for $G_n^{xy}(t)$ on the form (again suppressing the time argument),
\begin{equation} \label{Gn}
G_n^{xy} = -\frac{i}{4} \langle \phi_1 \phi_2 \phi_3 \ldots \phi_{2n} \rangle ,
\end{equation}
where $\langle \phi_1 \phi_2 \phi_3 \ldots \phi_{2n} \rangle$ can be written as the Pfaffian of a skew-symmetric matrix [\onlinecite{Fubini}], 
\begin{eqnarray} \label{pf}
 \mbox{pf} \left( {\begin{array}{*{20}{c}}
{\left\langle {{\phi _1}{\phi _2}} \right\rangle }&{\left\langle {{\phi _1}{\phi _3}} \right\rangle }&{\left\langle {{\phi _1}{\phi _4}} \right\rangle }& \cdots &{\left\langle {{\phi _1}{\phi _{2n}}} \right\rangle }\\
{}&{\left\langle {{\phi _2}{\phi _3}} \right\rangle }&{\left\langle {{\phi _2}{\phi _4}} \right\rangle }& \cdots &{\left\langle {{\phi _2}{\phi _{2n}}} \right\rangle }\\
{}&{}&{\left\langle {{\phi _3}{\phi _4}} \right\rangle }& \cdots &{\left\langle {{\phi _3}{\phi _{2n}}} \right\rangle }\\
{}&{}&{}& \ddots & \vdots \\
{}&{}&{}&{}&{\left\langle {{\phi _{2n - 1}}{\phi _{2n}}} \right\rangle }
\end{array}} \right),
\end{eqnarray}
with the operators $\phi_j, j=1,2, \ldots 2n$, identified from 
\begin{multline}
\langle \phi_1 \phi_2 \phi_3 \ldots \phi_{2n-2} \phi_{2n-1} \phi_{2n} \rangle \\ 
= \langle B_1 A_2 B_2 A_3 \ldots B_{n-1} A_{n} B_{n} B_{n+1} \rangle.
\end{multline}
Carrying out a Bogoliubov transformation to the post-quench diagonal basis and time evolving the matrix elements one obtains 
closed expressions for the matrix elements in (\ref{pf}); see Eqs. (A15) - A(17)  in the Appendix. 

The calculation of the other two-point functions which appear in Eq. (\ref{xiclosed}) proceed analogously, with details accounted for in the Appendix. Assembling the results and calculating the Pfaffians numerically we finally obtain the time evolution of spin squeezing after a quench $h_1 \rightarrow h_2$, represented by a plot of $\xi_s^2(t)$ given the pre-quench ground state of the XY chain for $h=h_1$. To this we turn next.}

\begin{figure}
\centerline{\includegraphics[width=0.9\linewidth]{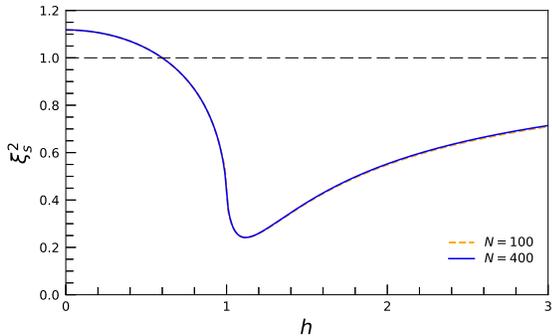}}
\caption{(color online). Equilibrium zero-temperature behavior of the spin-squeezing parameter of the XY chain as a function of the transverse field $h$ when $\delta=0.8$ for chain sizes $N=100, 400$. The intersection with the black dashed line, $\xi_s^2=1.0$, corresponds to a spin-coherent state at $h=0.6$, with unsqueezed (squeezed) spin states for $h<0.6 \ (h>0.6)$.}
\label{Fig1}
\end{figure}
\section{Time evolution of spin squeezing after a quench}

As a backdrop to our results for the post-quench behavior of the spin-squeezing parameter, let us first look at its dependence on the transverse field $h$ in the ground state of the model, i.e., {\em before} a quench. {\color{black} From Eq. (\ref{xiclosed}), using the results in the Appendix to implement the computation of the pre-quench spin correlation functions at $t=0$}, we find that the ground state phase diagram separates into two regions: one fully within the FM phase and bounded by $h<\sqrt{1-\delta^2}$ where there is no spin squeezing $(\xi^2_s > 1)$, with the other, complementary region $h>\sqrt{1-\delta^2}$, featuring spin squeezing {\color{black} $(\xi^2_s < 1)$}. The boundary between the two regions, $h=\sqrt{1-\delta^2}$, supports spin coherence $(\xi^2_s = 1)$. In Fig. \ref{Fig1} we illustrate the situation by plotting $\xi^2_s$ as a function of $h$, choosing $\delta = 0.8$. As one tunes $\delta$, the magnitude of the transverse field for which the spin squeezing is at its largest (smallest $\xi^2_s$) gets shifted from $h \approx 1.14$ when $\delta =1$ (transverse field Ising model) to the quantum critical point $h_c=1$ when $\delta \rightarrow 0$ (isotropic XX limit). As transpires from Fig. \ref{Fig1}, the 
pre-quench equilibrium squeezing parameter is effectively size-independent, in agreement with the finding in Ref. [\onlinecite{Liu}]. While the same holds true also for the squeezing parameter of the post-quench nonequilibrium states at short and intermediate time scales [\onlinecite{HadiNumerical}], its size dependence will grow with time and eventually become visible. We shall explore this latter phenomenon in Sec. V, but for now, in this section, we choose to work with a fixed system size, in the subsequent numerical plots taken to be {\color{black} $N=100$}.  

In this context, and in the light of works which have exploited spin squeezing as a means to identify quantum entangled states [\onlinecite{Wang,Toth,Guehne}], let us point out that the spin-squeezing parameter here fails to detect the ground state entanglement. This is so since it is known that there is no qualitative change of the entanglement across the boundary $h=\sqrt{1-\delta^2}$ in this model, not for the two-spin entanglement of formation [\onlinecite{Osterloh,Osborne}] and also not for the entanglement entropy for two blocks [\onlinecite{Vidal}].   

In the following we explore the time-dependent behavior of the squeezing parameter after a quench from squeezed, unsqueezed, and coherent ground states of the XY model. Most interestingly, we find that we can {\color{black} achieve squeezed} states $-$ also at large times $-$ from initial states that are neither coherent nor squeezed. {\color{black} 

We also present results for the time-dependent variance of the mean spin direction, $\nu(J_z(t)) = {\color{black} \langle J_z^2\rangle - \langle J_z\rangle^2}$, intriguingly showing a dynamical covariation with the spin-squeezing parameter for some quench scenarios, but not for others.} {\color{black} It can be calculated from
\begin{equation} \label{msd}
\nu(J_z(t)) = N\big(\frac{1}{4} + \sum_{n=1}^{N-1} G_n^{zz}(t)\big) - \big( NM_z(t)\big)^2,
\end{equation} 
with $M_z(t) = \langle J_z \rangle /N$ the time-dependent magnetization.} 

{\color{black}To set the stage, we shall begin to study the case when the initial state is squeezed. This allows us to cover all three possible types of initial states of the model $-$ FM, critical, and PM $-$ providing a helpful context to the more interesting cases when the initial states are unsqueezed or coherent (allowing only for a ferromagnetic initial state; cf. Fig.1).}

\subsection{ Squeezed initial state }
In this first section we consider altogether fourteen different quench scenarios, taking off from three distinct squeezed initial states, all being ground states of a (pre-quench) XY Hamiltonian: (I) one in the FM phase, at $h_1=0.8$;, (II) one exactly at the quantum critical point, $h_1=1.0$; and (III) one in the PM phase, at $h_1=2.0$. For each of these initial states we then perform quenches by choosing a post-quench XY Hamiltonian with a ground state which is (a) unsqueezed; (b) spin coherent; (c) squeezed and in the same phase as the initial state; and (d) squeezed and in the other phase than the one to which the initial state belongs. In addition, we consider (IV) a quench from a squeezed FM and PM phase, respectively, to the equilibrium quantum critical point {\color{black} (i.e., with the ground state of the post-quench XY Hamiltonian being squeezed {\em and} critical). Note that while the ground state of a post-quench Hamiltonian is an equilibrium state}, the actual time-dependent {\em (post-quench)} states are nonequilibrium states, expected to equilibrate to a (mixed) GGE state in the thermodynamic limit at large times [\onlinecite{Essler}]. While it should be obvious that {\color{black} ground states,} post-quench states, and GGE states have very different characters, we point this out only to avoid confusion when referring to the various quench scenarios above. 
 
\begin{figure}[t]
\centerline{
\includegraphics[width=1\linewidth]{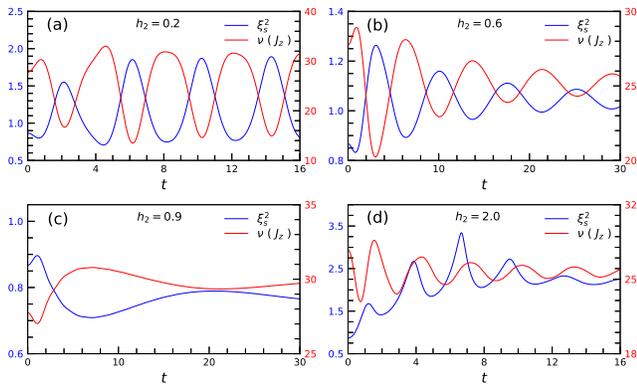}}
\caption{(color online). Time evolution of the spin-squeezing parameter {\color{black} $\xi_s^2$ (blue) and the variance of the mean spin direction $\nu(J_z(t))$ (red)} at $\delta=0.8$ for quenches from {\color{black} the FM state at} $h_1=0.8$ to (a) $h_2=0.2$, (b) $h_2=0.6$, (c) $h_2=0.9$, and (d) $h_2=2.0$.}
\label{Fig2}
\end{figure}

\begin{figure}
\centerline{
\includegraphics[width=1\linewidth]{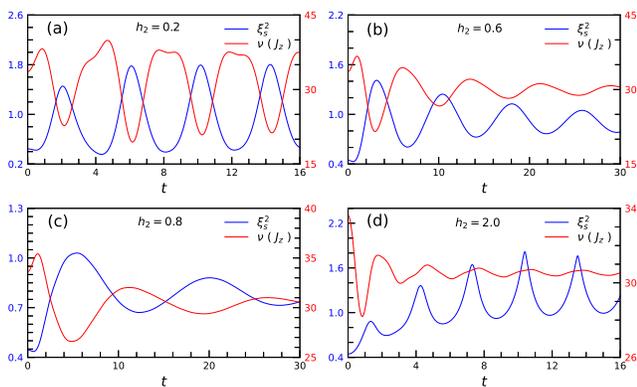}}
\caption{(color online). Time evolution of {\color{black} $\xi_s^2$ (blue) and $\nu(J_z(t))$ (red)} at $\delta=0.8$  for quenches  from {\color{black} the critical state at} $h_1=h_c=1.0$ to (a) $h_2=0.2$, (b) $h_2=0.6$, (c) $h_2=0.8$, and (d) $h_2=2.0$.}
\label{Fig3}
\end{figure}

\begin{figure}
\centerline{
\includegraphics[width=1\linewidth]{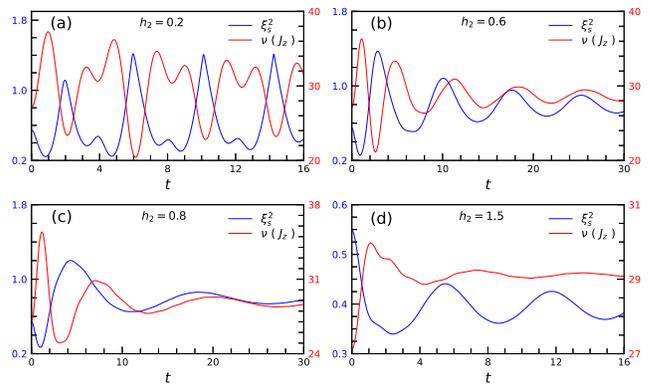}
}
\caption{(color online). Time evolution of {\color{black} $\xi_s^2$ (blue) and $\nu(J_z(t))$ (red)} at $\delta=0.8$  for quenches from {\color{black} the PM state at} $h_1=2.0$  to (a) $h_2=0.2$, (b) $h_2=0.6$, (c) $h_2=0.8$ and (d) $h_2=1.5$.}
\label{Fig4}
\end{figure}

\begin{figure}
\centerline{
\includegraphics[width=1\linewidth]{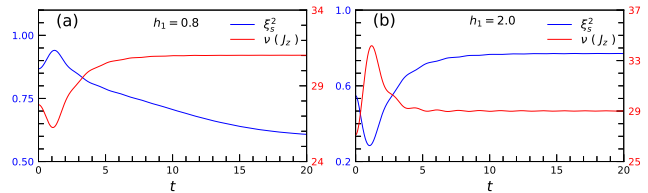}
}
\caption{(color online). Time evolution of {\color{black} $\xi_s^2$ (blue) and $\nu(J_z(t))$ (red)} at $\delta=0.8$  for quenches from (a) $h_1=0.8$ {\color{black} (FM)} and (b) $h_1=2.0$ {\color{black} (PM)} to $h_2=h_c=1.0$.}
\label{Fig5}
\end{figure}

The time dependence of the spin-squeezing parameter $\xi^2_s(t)$ for the various quench scenarios is plotted in Figs.~\ref{Fig2}-\ref{Fig4}, with the panels (a)-(d) corresponding to the protocols (a)-(d) above, and Fig.~\ref{Fig5} with panels (a) and (b) belonging to quenching into the critical point. In all cases, the anisotropy parameter has been set to $\delta=0.8$, implying a spin-coherent state at $h=0.6$, with unsqueezed (squeezed) spin states for $h\!<\!0.6 \ (h\!>\!0.6)$. The horizontal and vertical black dashed lines in the plots mark $\xi_s^2=1.0$ (boundary between squeezed and unsqueezed states and corresponding to a spin-coherent state) and $h_c=1.0$ (quantum critical point between FM and PM phases), respectively. Let us now walk through the different cases.

\subsubsection*{(I) Ferromagnetic initial state, Fig.~\ref{Fig2}
}   
\vspace{-0.35cm}
A notable feature in the time evolution of the squeezing parameter $\xi^2_s(t)$, illustrated in Fig.~\ref{Fig2} (blue color), is its oscillation, manifest beyond an early transient time and most easily visible in panels (a), (b), and (d): $\xi^2_s(t)$ oscillates around a mean value with an amplitude which decreases with time $-$ faster or slower depending on the type of quench $-$ and with a quench-dependent period $T_\text{osc}$. 

It is also interesting to note {\color{black} the increase of $\xi_s^2(t)$ at short time scales after the quench in panels (a), (b), and (d). One may think of} this as caused by the modes suddenly excited by the quench, which scramble the quantum fluctuations so that the spin squeezing decreases. Depending on the particular protocol, squeezing may then be permanently lost, (Fig.~\ref{Fig2}(d), or recovered only transiently, Figs.~\ref{Fig2}(a),(b). Only if the {\color{black} initial state and the ground state of the post-quench Hamiltonian} are both squeezed {\em and} belong to the FM phase do the post-quench states remain squeezed for all times, (Fig.~\ref{Fig2}(c)).    

{\color{black} Turning to the variance in the mean spin direction, $\nu(J_z(t))$ (red color in Fig. 2), one notes a perfect antiphase variation with respect to the spin-squeezing parameter in panels (a)-(c), but not in (d). As we shall see, this loss of covariation is generic when quenching across the quantum critical point, from the FM to the PM phase or vice versa.} 

\subsubsection*{(II) Quantum critical initial state, Fig.~\ref{Fig3}} 
\vspace{-0.35cm}
{\color{black} Features from case (I) show up also when the initial state is critical: the oscillations of $\xi^2_s(t)$, and an increase of $\xi_s^2(t)$ at short time scales} after the quench. Different from the previous case (I), none of the panels in Fig.~3 display a covariation between $\xi_s^2(t)$ and $\nu(J_z(t))$. As {\color{black} indicated by} numerical results from other choices of quench parameters (not shown here), one {\color{black} is led to infer} that a quench {\em from} the quantum critical point {\em always} corrupts the covariation [\onlinecite{HadiNumerical}].   

\begin{figure*}[t]
\centerline{
\includegraphics[width=1\linewidth]{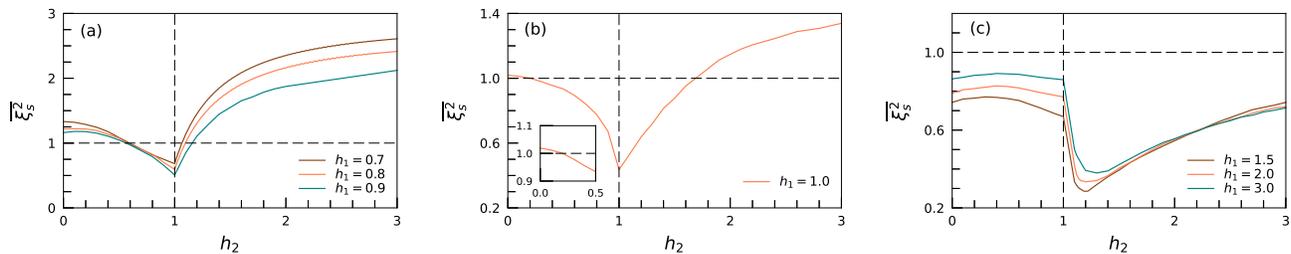}
}
\caption{(color online).  Long-time average {\color{black} $\overline{\xi_s^2}$} of the spin-squeezing parameter versus the final quenched field $h_2$  when $\delta=0.8$, for quenches from (a)  $h_1=0.7, 0.8, 0.9$ {\color{black} (FM)}, (b) $h_1=h_c=1.0$ (critical) and (c) $h_1=1.5, 2.0, 3.0$ {\color{black} (PM)}.  
}
\label{Fig6}
\end{figure*}

\subsubsection*{ (III) Paramagnetic initial state, Fig.~\ref{Fig4}} 
\vspace{-0.35cm}
The overall structure of the quench dynamics is recognizable from cases (I) and (II): {\color{black} With the exception of panel (d) of Fig. 4, the spin squeezing decreases at short time scales (barring a transient behavior right after the quench). Further, the spin squeezing parameter $\xi^2_s(t)$ exhibits an oscillating behavior in all panels (a) - (d). With the exception of panel (a), there is no distinct covariation between $\xi_s^2(t)$ and $\nu(J_z(t))$.} As evidenced by other choices of parameter values (not shown here), this is generic for any quench {\em from} the PM phase {\color{black} to the unsqueezed region in the FM phase [\onlinecite{HadiNumerical}].}

\subsubsection*{ (IV) {\color{black} Quench to the quantum critical point}, Fig.~\ref{Fig5}}
\vspace{-0.35cm}
The two cases displayed in Figs.~\ref{Fig5}(a) (FM initial state) and \ref{Fig5}(b) (PM initial state) are markedly different from the previous cases (I) - (III): {\color{black} The oscillating behavior in the time evolution of $\xi^2_s(t)$ seen in all quench scenarios of} (I) - (III) is absent. Instead, a steady-state spin squeezing {\color{black} becomes apparent at the larger} intermediate times displayed in the plots [i.e., before a revival eventually sets in; cf. Sec. V], with an enhanced (reduced) squeezing as compared to the initial FM (PM) state. {\color{black} Note also the covariation of $\nu(J_z(t))$ with $\xi^2_s(t)$ in both panels.} 

\subsubsection*{Long-time average of the spin-squeezing parameter, Fig.~\ref{Fig6}}
\vspace{-0.35cm}
To obtain a deeper insight into the quench dynamics of the system it is instructive to compute the time average $\overline{\xi _s^{2}}$ of the spin-squeezing parameter, 
\begin{eqnarray}\label{eq13}
\overline{\xi _s^{2}} := \lim_{T \rightarrow \infty} \frac{1}{T} \int_0^T \xi _s^2(t)~ dt. 
\label{tangle}
\end{eqnarray}
In Fig.~\ref{Fig6} we have plotted $\overline{\xi _s^{2}}$ as a function of the transverse field $h_2$ in the post-quench Hamiltonian, with initial squeezed states (a) in the FM phase, (b) at the quantum critical point, and (c) in the PM phase. Figs.~\ref{Fig6}(a),(b) reveal that by fine tuning $h_2$ to one of two possible values one may achieve a spin-coherent state in the long-time equilibrium state, provided that the initial state is not in the PM phase. In contrast, if the initial state {\em is} in the PM phase, {\color{black} Fig.~\ref{Fig6}(c)}, the spin will always be squeezed at large times, independent of the choice of $h_2$.

Most strikingly, all quenches performed onto the quantum critical point, $h_2=1.0$, lead to a squeezed state with a nonanalyticity in $\overline{\xi _s^{2}}$, suggestive of a nonequilibrium phase transition with $h_2$ as control parameter. This is somewhat reminiscent of results for the transverse field Ising chain with long-range interaction where it has been argued that the long-time average of the magnetization may disclose nonequilibrium criticality when quenching across an equilibrium quantum critical point [\onlinecite{Knap}]. For related results, see Refs.~[\onlinecite{Zhou,Hadi2,Hadi3}]. Note, however, that in the present case a nonanalyticity appears only when the {\color{black} ground state of the post-quench Hamiltonian} itself is critical. 

 \subsection{Unsqueezed initial state}
 
\begin{figure}[t]
\centerline{\includegraphics[width=1\linewidth]{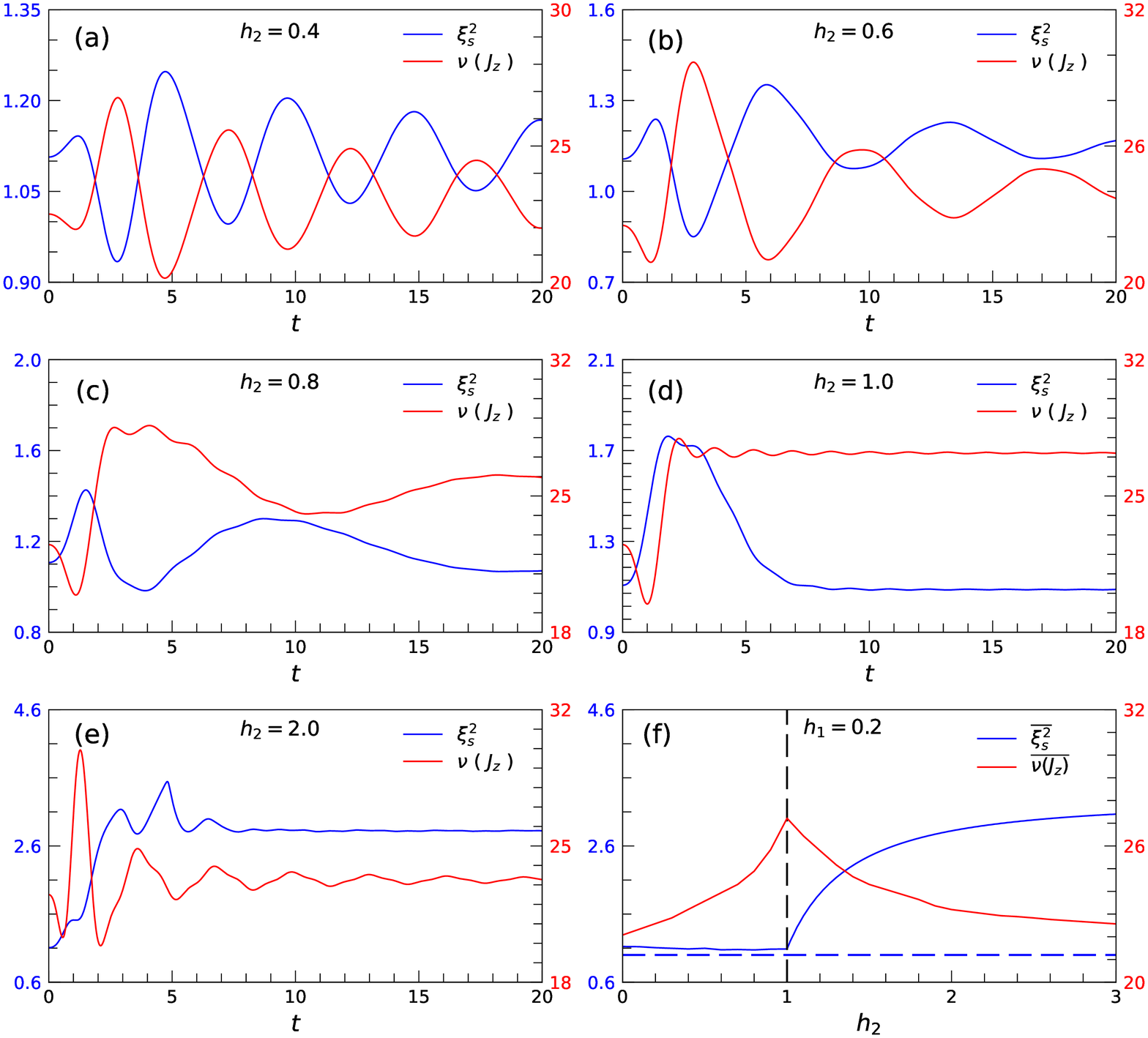}}
\caption{(color online). Time evolution of the spin-squeezing parameter {\color{black} $\xi_s^2$ (blue) and the variance of the mean-spin direction $\nu(J_z(t))$ (red)} when $\delta=0.8$ for the initial state at $h_1=0.2$, with a quench into (a) an unsqueezed state ($h_2=0.4$); (b) the coherent state ($h_2=0.6$); (c) a squeezed state at $h_2=0.8<h_c$; (d) a squeezed state at the quantum critical point $h_2=h_c=1.0$; and (e) a squeezed state at $h_2=2.0>h_c$. Panel (f) displays {\color{black} plots of the long-time averages $\overline{\xi_s^2}$ (blue) and  {\color{black} $\overline{\nu (J_z)}$} (red) versus $h_2$ of the spin-squeezing parameter and the variance of the mean spin direction, respectively,} for a quench started from $h_1=0.2$, also with $\delta=0.8$. {\color{black} The dotted blue line marks $\overline{\xi_s^2} = 1$.}}
\label{Fig7}
\end{figure}

\begin{figure}[t]
\centerline{\includegraphics[width=1\linewidth]{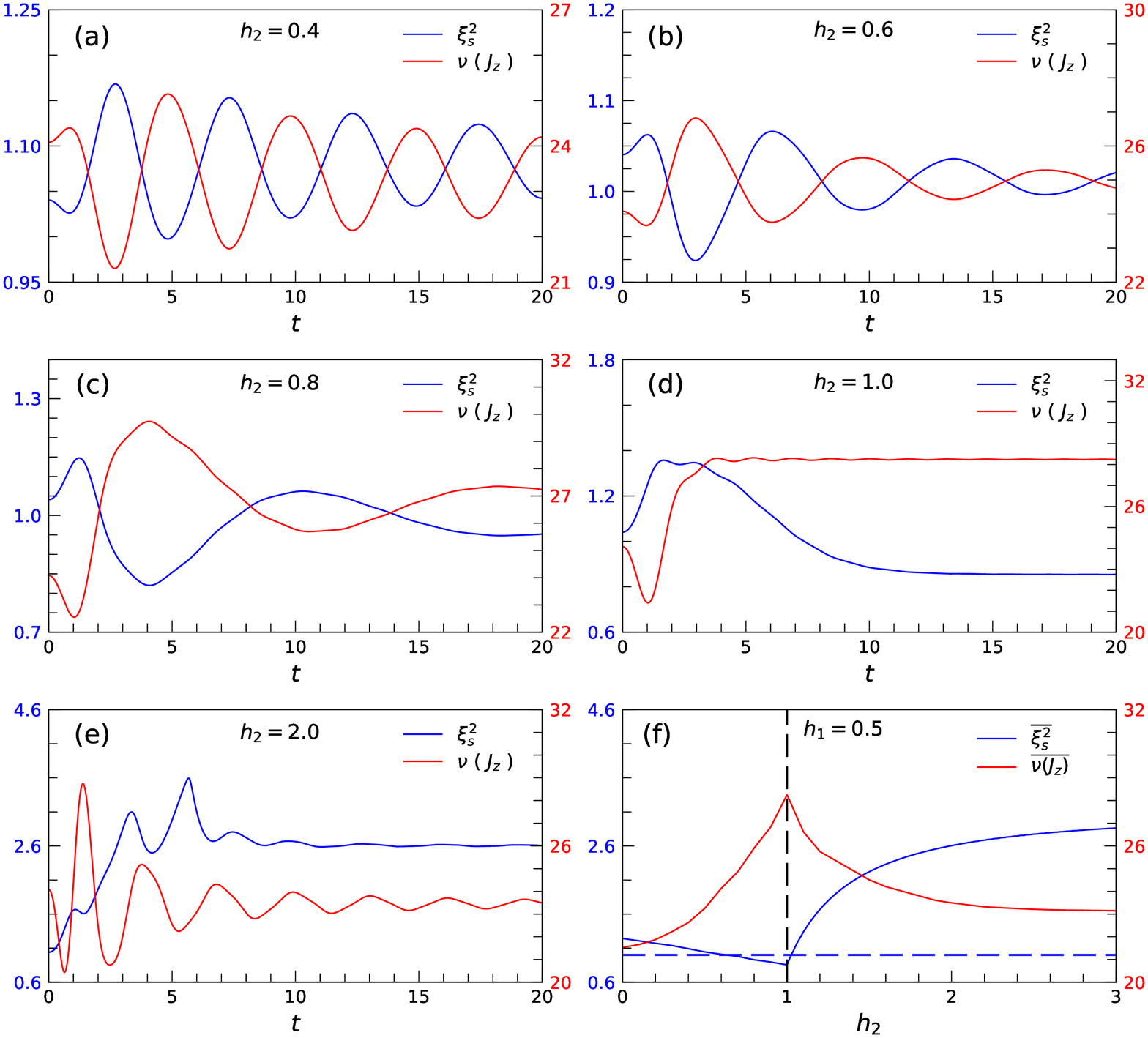}}
\caption{(color online). {\color{black} Same as Fig.~\ref{Fig7} but with initial state at $h_1=0.5$.}}
\label{Fig8}
\end{figure}

In this section we consider ten types of quench scenarios, taking off from one of two unsqueezed initial states {\color{black} (always in the FM phase; cf. Fig. 1)}: the ground states of a pre-quench XY Hamiltonian with $\delta = 0.8$ (as before), {\color{black} now choosing $h_1=0.2$ (Fig.~\ref{Fig7}) and $h_1=0.5$ (Fig.~\ref{Fig8})}, respectively. The five {\color{black} types of ground states of the post-quench Hamiltonian} (also with $\delta=0.8$ but with different transverse fields $h_2$) are chosen as (a) an unsqueezed state at $h_2=0.4$; (b) a coherent state, i.e., a state at $h_2=0.6$; (c) a squeezed state in the FM phase at $h_2=0.8$; (d) the squeezed state at the quantum critical point $h_2=1.0$; and (e) a squeezed state in the PM phase at $h_2=2.0$. The time dependencies of the spin-squeezing parameter $\xi^2_s(t)$ {\color{black} and the variance of the mean spin direction $\nu(J_z(t))$} for the different quench scenarios are plotted in Fig.~\ref{Fig7}(a)-(d) and \ref{Fig8}(a)-(d), with the panels labeled according to the protocols (a)-(d) above. 

Features from the plots in Figs.~\ref{Fig2}-\ref{Fig5} where the initial state was squeezed are also present in several of the panels of Figs.~\ref{Fig7} and \ref{Fig8}: {\color{black} the increase of $\xi^2_s(t)$ on short time scales} after the quench (panels (b)-(e)); and oscillations of {\color{black} $\xi^2_s(t)$} with a quench-dependent period (panels (a) and (b)). {\color{black} As for the variance of the mean spin direction, $\nu(J_z(t))$, one observes a near-perfect anti-phase variation with $\xi_s^2(t)$ for the quenches in panels (a) and (b), with a weakly perturbed covariation {\color{black} seen} in (c) and (d).}

The expectation that a quench from an initially unsqueezed state can produce only unsqueezed post-quench states is met in all cases unless the transverse field in the pre-quench Hamiltonian is chosen not too far from that which results in spin coherence, e.g., $h_1=0.5$ as in Fig.~\ref{Fig8}. With that choice, a quench to the same FM phase, Fig.~\ref{Fig8}(c), or to the quantum critical point, Fig.~\ref{Fig8}(d), will yield a {\color{black} {\em squeezed} post-quench state. Moreover, in the latter case the squeezed state is quasi-stationary (i.e., a state where the spin squeezing is approximately constant at intermediate time scales $-$ here with extremely small fluctuations not captured by the plots $-$ before a quantum revival sets in at later times; cf. Sec. V).} This surprising outcome of quench dynamics is also strikingly seen in the time average $\overline{\xi_s^2}$, plotted in Fig.~\ref{Fig8}(f). In addition, this figure also illustrates that one may produce, in the mean, a coherent spin state from an unsqueezed state by a judicious choice of the quench parameter, i.e., by adjusting the value of $h_2$. Taken together, this shows that a quantum quench may redistribute quantum fluctuations in such a way as to produce spin coherence or spin squeezing from an initially unsqueezed state. This surprising effect should in principle be controllable in an experiment as it is induced by tuning a magnetic field.

\begin{figure}[t]
\centerline{\includegraphics[width=1\linewidth]{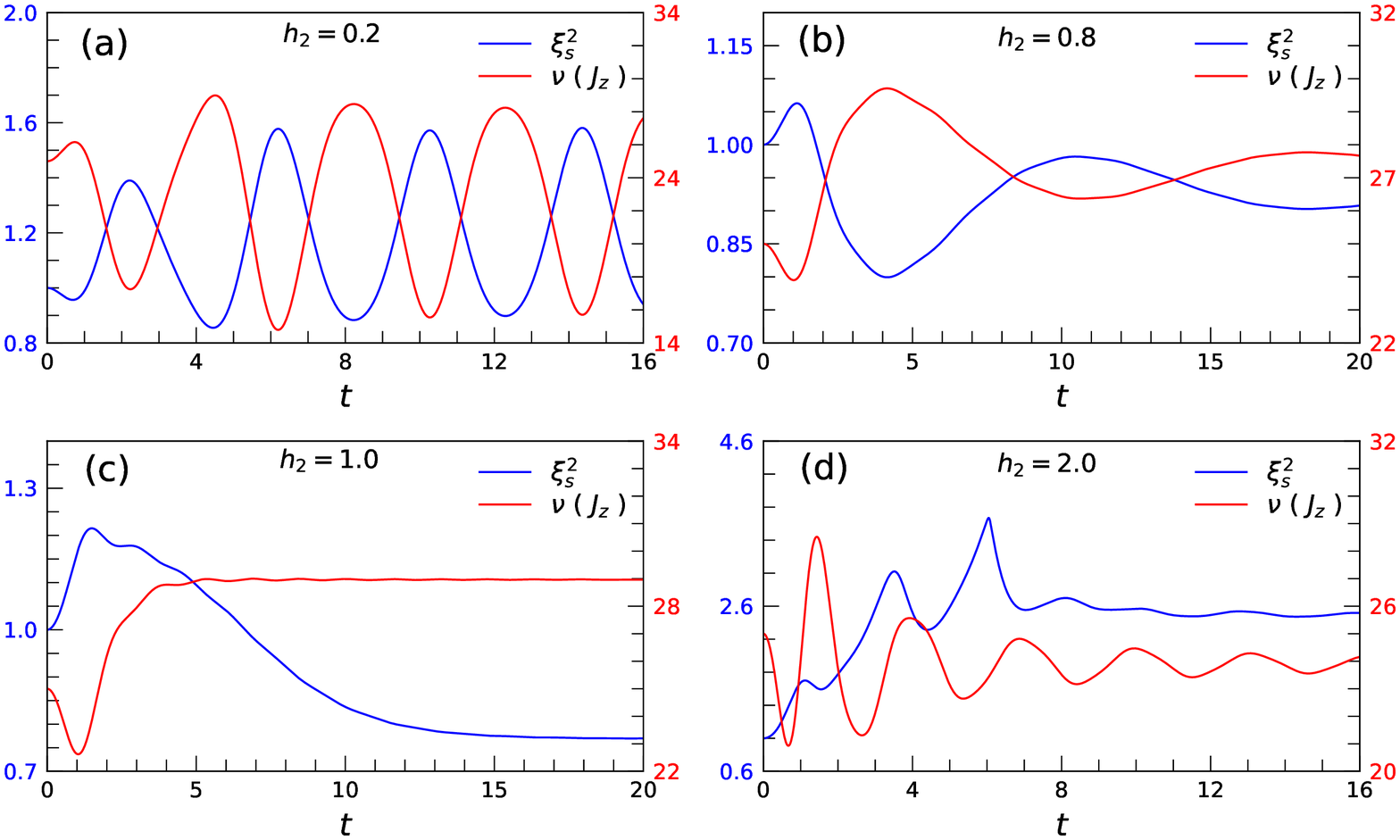}}
\centerline{\includegraphics[width=0.5\linewidth]{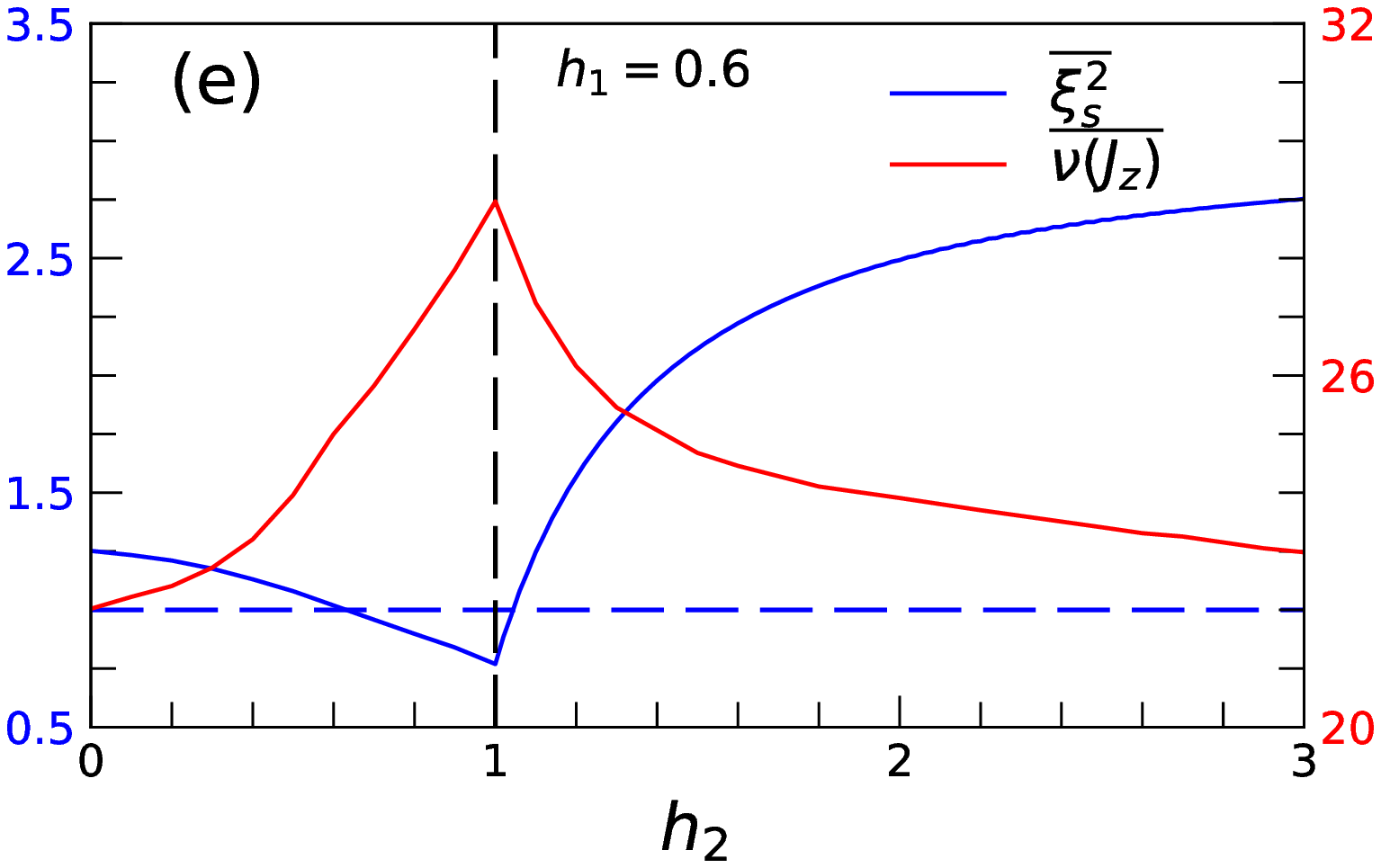}}
\caption{(color online). Time evolution of the spin-squeezing parameter {\color{black} $\xi_s^2$ (blue) and the variance of the mean-spin direction $\nu(J_z(t))$ (red)} when $\delta=0.8$ for quenches from the coherent state at $h_1=0.6$  to (a) $h_2=0.2$, (b) $h_2=0.8$, (c) $h_2=h_c=1.0$ and (d) $h_2=2.0$. Panel (e) displays {\color{black} plots of the long-time averages {\color{black} $\overline{\xi_s^2}$} (blue) and {\color{black} $\overline{\nu (J_z)}$} (red) versus $h_2$ of the spin-squeezing parameter and the variance of the mean spin direction, respectively,} for a quench started from $h_1=0.6$. {\color{black} The dotted blue line marks $\overline{\xi_s^2} = 1$.}} 
\label{Fig9}
\end{figure}

 \subsection{ Coherent initial state}
A spin-coherent state is defined to be a minimum-uncertainty state where the variance of the many-particle spin operator is equal in all directions [\onlinecite{Radcliffe}]. It is often described as a state which has dynamics that closely resembles that of a classical harmonic oscillator. Accordingly, it can appear as a ground state factorized into a direct product of single spin states. For the Hamiltonian (\ref{eq1}), the factorization is realized at  $h=\sqrt{1-\delta^2}$ [\onlinecite{Adesso}] so that for $\delta=0.8$, its value will be $h=0.6$, a result that we have used repeatedly in Secs. IV.A and B. The results for the time dependence of the spin-squeezing parameter after a quench from an initial ground state of the XY chain at $h_1=0.6$ are illustrated in Fig.~\ref{Fig9}. We here consider quench scenarios with four different {\color{black} post-quench fields: (a) $h_2=0.2$ (FM unsqueezed ground state of ${\cal H}(h_2)$); (b) $h_2=0.8$ (FM squeezed ground state of ${\cal H}(h_2)$); (c) $h_2=h_c=1.0$ (critical point); and (d) $h_2=2.0$ (PM squeezed ground state of ${\cal H}(h_2)$)} .  

\begin{figure*}[t]
\centerline{
\includegraphics[width=1\linewidth]{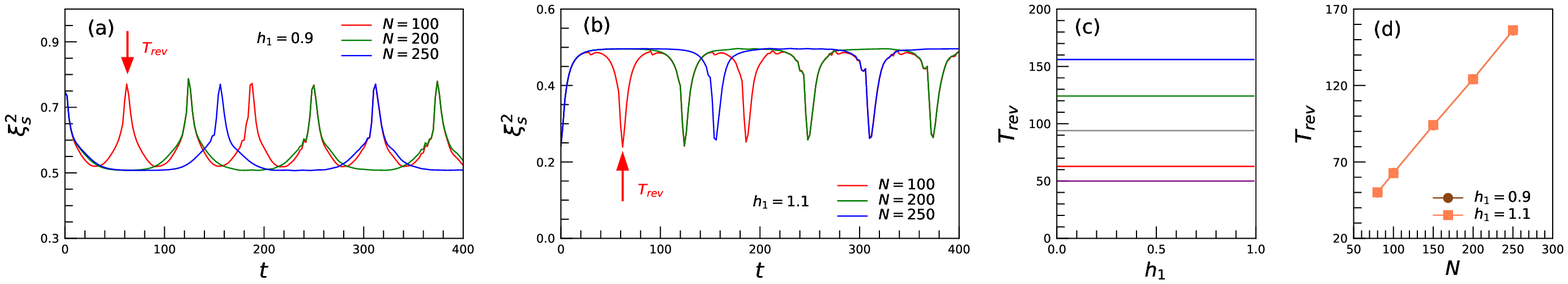}
}
\caption{(color online). 
Long-time dynamics of the spin-squeezing parameter for $\delta=0.8$ for quenches into $h_2=h_c=1.0$ from (a) $h_1=0.9$ and (b) $h_1=1.1$ for system sizes $N=100, 200, 250$. The first revivals for $N=100$ are marked by arrows. Panel (c) displays the revival time $T_{rev}$ vs. $h_1$ for quenches from the FM region $0 \le h_1<1$ to the critical point $h_2 = 1.0$ for system sizes $N=80, 100, 150, 200, 250$ (from purple to blue), showing that $T_{rev}$ is insensitive to the choice of initial ferromagnetic state. Panel (d) exhibits the linear scaling of the revival time with system size, $T_{rev} = kN$. To within numerical accuracy, $k=0.625 \pm 0.005$, consistent with the expected revival structure (cf. main text).}
\label{Fig10}
\end{figure*}

While a quench to $h_2$ is seen to induce coherent (squeezed) states at repeated times (time intervals), cf. Fig.~\ref{Fig9}(a), the quench to $h_2=2.0$ in Fig.~\ref{Fig9}(d) never returns a coherent or squeezed state. Differently, and disregarding a short-time transient behavior, a quench to $h_2=0.8$, {\em or} to the critical point, produces squeezed post-quench states at all times, cf. Figs.~\ref{Fig9}(b) and \ref{Fig9}(c).  {\color{black} As expected from Fig.~\ref{Fig8}(d) $-$ corresponding to a quench from a moderately unsqueezed state to the critical point $-$ as well as from the ``standard spin-squeezing protocol" (where the system is initialized in a coherent spin state [\onlinecite{Radcliffe}]), in quench scenario (c) the spin squeezing becomes quasi-stationary (cf. the discussion in the previous subsection).} 

The time average $\overline{\xi^2_s}$ in the thermodynamic limit as a function of the quench field $h_2$ is shown in Fig.~\ref{Fig9}(e). Most interestingly, the plot of 
$\overline{\xi^2_s}$ reveals that there is a region of values for the post-quench field $h_2$, in an interval around the critical field $h_c = 1.0$,  yielding spin-squeezed states in the long-time limit. As is also seen in the figure, by fine-tuning to a particular value of $h_2$, a coherent quasi-stationary state can also be produced on the average.  

As in previous quench scenarios $-$ starting from squeezed or unsqueezed ground states $-$ the nonanalyticity of $\overline{\xi^2_s}$ at the equilibrium quantum critical point $h_2=1.0$ again suggests {\color{black} that the ferro- and paramagnetic phases also define distinct phases for how the transverse field redistributes  quantum fluctuations among the spin components away from equilibrium.} No such behavior is seen in {\color{black} the long-time average $\overline{\nu (J_z)}$}. Turning to the time dependence of $\nu(J_z(t))$, one notes a covariation (Fig.~\ref{Fig9}(a),(b)), or a weakly perturbed covariation ((Fig.~\ref{Fig9}(c)), with $\xi_s^2(t)$. 

\section{Universality of the revival  times}

A useful concept in the study of quantum quenches is that of the {\em Loschmidt echo} [\onlinecite{Gorin}] which measures the overlap between pre-quench and post-quench states. Numerical finite-size studies have revealed that the time dependence of the Loschmidt echo of several models exhibits a periodic pattern after a quantum critical quench $-$ a ``revival structure" $-$ formed by sudden large deviations from its mean value when {\color{black} quenching} to a quantum critical point [\onlinecite{Quan,Yuan,Rossini,Zhong}], i.e., in the language of the XY chain, choosing $h_2=h_c= 1.0$. The amplitudes of these revivals may decay with time; however, it has been argued that under certain conditions the revival structure may be periodic in time, with a universal expression, $T_\text{rev} = N/2v_g$, for the period $T_\text{rev}$ which is independent of the initial state and the size of the quench [\onlinecite{Hamma,Henrik,Delgado}]. Here $N$ is the number of sites of the periodic lattice, with $v_g$ generically being the magnitude of the maximum group velocity of quasiparticles of the post-quench Hamiltonian. However, there are exceptions [\onlinecite{Fago,Dubail,Najafi}]. For example, when the post-quench XY Hamiltonian is tuned to criticality ($h_2=1.0$), $v_g=\delta < v_{g, \text{max}}$ [\onlinecite{Najafi}]. Analogous results have been found also for revivals of the maximal quantum Fisher information in the transverse field Ising chain (corresponding to $\delta = 1$ in the XY chain) [\onlinecite{Hadi2}]. 

Could it be that the spin-squeezing parameter $\xi_s^2(t)$ in the present model also exhibits a revival structure for large times (as compared to the short and intermediate times examined in Sec. IV)? If so, will its revival period also follow the universal expression $T_\text{rev} = N/2v_g$?  

Here, focusing on the case where the {\color{black} ground state of the post-quench Hamiltonian} is critical, i.e., with $h_2=1.0$, the answer is yes. Figure~\ref{Fig10} exhibits the time dependence of $\xi^2_s(t)$ over large time intervals for quenches from a ferromagnetic (paramagnetic) squeezed ground state of the XY chain with $\delta=0.8$ at $h_1=0.9$ ($h_1=1.1$) to the {\color{black} equilibrium quantum critical point}, $h_2=1.0$.  Results are shown for different system sizes {\color{black}$N= 80, 100, 150, 200, 250$}. Inspection of Fig.~\ref{Fig10}(d) shows that the spin-squeezing dynamics manifests the same universal revival structure as that of the Loschmidt echo [\onlinecite{Hamma,Henrik}], with a linear scaling of the revival time $T_{rev}$ with system size $N$. The numerical data yield a slope {\color{black}$0.625 \pm 0.005$}, implying that the formula $T_{rev} = N/2v_g$ predicts that $v_g \approx 0.8 = \delta$, in excellent agreement with Ref. [\onlinecite{Najafi}]. 


\section{Summary}

Quantum technologies currently receive an enormous amount of attention because of their potential beyond classical limits.
Spin squeezing is one of the most promising strategies for attaining a quantum advantage in practical sensing applications. The idea of squeezing aims to reduce quantum uncertainties intrinsic no noncommuting observables, and by that, to achieve high-precision measurements. A question that is only now beginning to be addressed is how to understand the time-dependence of spin squeezing in the nonequilibrium dynamics of a quantum many-body system [\onlinecite{Hazzard,Perlin,Lerose,Xu2}]. 

{\color{black} In this paper} we have tried to add to this line of research by studying the dynamical spin squeezing after sudden quenches in the spin-1/2 XY chain with a transverse magnetic field.   
This model is exactly solvable [\onlinecite{FranchiniBook}], allowing access to {\color{black} well-controlled} results, and also, making our study of interest for understanding nonequilibrium phenomena of integrable systems. 

We have found that the post-quench nonequilibrium states in a finite system typically fluctuate between squeezed, coherent, and unsqueezed states at short time scales $-$ as measured by the spin-squeezing parameter [\onlinecite{Kitagawa}] $-$ and may then settle for one of these types at intermediate times before a revival sets in. Intriguingly, a proper choice of the quench protocol makes it possible to produce squeezed spin states starting from a ground state of the XY chain which is neither coherent nor squeezed, the sole requirement being that the magnitude of the applied transverse magnetic field lies within a certain interval. {\color{black} This suggests an alternative to the standard spin-squeezing protocol $-$ exploited in most experiments [\onlinecite{Ma1}] $-$ where one initializes a system in a coherent state and then applies a Hamiltonian which shears the spin distribution. To answer the question for what specific experimental setups, applications, or tasks it may prove advantageous to forego a spin-coherent initial state $-$ a viable option according to our finding! $-$ requires more work, both theoretical and experimental. Let us here point to a recently proposed quantum simulator of 1D spin-exchange models $-$ including the XY chain in a transverse magnetic field $-$ making use of circular Rydberg atoms as a platform [\onlinecite{Brune}]. Its realization holds great promise for future high-precision experimental tests of our predictions, including that of a covariation between spin squeezing and the variance of the mean spin direction in certain quench scenarios.} 

{\color{black} The time average $\overline{\xi^2_s}$ of the spin squeezing parameter features a non-analyticity} when quenching to the equilibrium quantum critical point of the model.  {\color{black} This} resembles the situation seen for other spin models where the long-time average of the magnetization discloses a nonequilibrium quantum phase transition when quenching across an equilibrium quantum critical point [\onlinecite{Knap,Zhou,Hadi2}]. It would be interesting to uncover the properties of this putative nonequilibrium transition, in the present case signaled by a cusp in $\overline{\xi^2_s}$ as a function of the quench parameter $h$. {\color{black} Another interesting problem is to understand how the properties of coherence and spin squeezing are imprinted in the limit of very large times [\onlinecite{Mallesh}], where, in the thermodynamic limit, the post-quench states are expected to equilibrate to a mixed state governed by a generalized Gibbs ensemble [\onlinecite{Essler}].} 

In conclusion, more theoretical work is certainly needed to fully understand the underlying physics of the various quench scenarios uncovered in this work. Their realizations would be fascinating, opening an experimental window into the use of quench dynamics as a tool for producing {\color{black} squeezed spin states, different from existing paradigms.}
\section*{Acknowledgments}
The authors thank Mansour Eslami, Mehdi Abdi, and Foroud Bemani for their good comments. {\color{black} We are also grateful to Tommaso Comparin and Tommaso Roscilde for pointing out an error in an early version of this work.} This research was done in a collaboration between the University of Guilan and the University of Gothenburg under Grant No. 178305/15. In addition, it was supported by the Swedish Research Council under Grant No. 621-2014-5972, and  the National Science Centre (NCN, Poland) under Grant No. 2019/35/B/ST3/03625.
\setcounter{equation}{0}
\renewcommand\theequation{A\arabic{equation}}
\section{Appendix}
{\color{black} To obtain the spin squeezing parameter $\xi_s^2(t)$ in Eq. (\ref{xiclosed}), and also the variance of the mean spin direction $\nu(J_z(t))$ in Eq. (\ref{msd}), we need expressions for the two-point functions
$G_n^{\alpha \beta} \!:=\! \langle {S_1^\alpha S_{1 + n}^\beta } \rangle, \alpha, \beta = x,y,z$. (Here, as in the main text, we have suppressed the time argument, using the notation $\langle \ldots \rangle := \langle \psi(t) | \ldots | \psi(t) \rangle$.)  Introducing $A_n = a_n^\dagger + a_n$ and $B_n = a_n^\dagger - a_n$, with $a_n^\dagger$ and $a_n$ the JW fermionic operators in Eq. (\ref{JWT}), a direct calculation using this same equation shows that
\begin{eqnarray}  \label{t1}
G_n^{xx } &\!=\!& \langle S_1^xS_{1 + n}^x \rangle  = \frac{1}{4}\langle B_1 A_2B_2...A_nB_nA_{n+1} \rangle    \nonumber \\
G_n^{yy} &\!=\!&\langle S_1^yS_{1+ n}^y \rangle  = \frac{( - 1)^n}{4}\langle A_1B_2A_2...B_nA_nB_{n+1} \rangle  \nonumber\\
G_n^{zz } &\!=\!&\langle S_1^zS_{1 + n}^z \rangle  = \frac{1}{4}\langle A_1B_1A_{n+1}B_{n+1} \rangle   
\end{eqnarray}
\begin{eqnarray}
G_n^{xy } &\!=\!&\langle S_1^xS_{1 + n}^y \rangle  = \frac{ - i}{4}\langle B_1A_2B_2...A_nB_nB_{n+1} \rangle \nonumber  \\
G_n^{yx } &\!=\!&\langle S_1^yS_{1 + n}^x \rangle  = \frac{  i( - 1)^n}{4}\langle A_1B_2A_2...B_nA_nA_{n+1} \rangle \nonumber \\. 
\end{eqnarray}
For $\alpha, \beta = x, y$, we can write these relations on the generic form
\begin{eqnarray} \label{eq2n}
G_n^{\alpha \beta}  = D_n^{\alpha \beta }\left\langle {{\phi _1}{\phi _2}{\phi _3}...\phi _{2n-2}\phi _{2n-1}{\phi _{2n}}} \right\rangle
\end{eqnarray}
with
\begin{eqnarray}
D_n^{xx} &\!=\!& \frac{1}{4}, \ \ \ D_n^{yy} = \frac{( - 1)^n}{4}, \nonumber \\
D_n^{xy} &\!=\!& \frac{ - i}{4}, \ \ \ D_n^{yx} \!=\! \frac{i( - 1)^n}{4},
\end{eqnarray}
and where each operator $\phi_j, j= 1,2, \ldots, 2n$, is identified with either an $A$ or $B$ operator with the proper index by a comparison with the corresponding expression in  (A1) or (A2).
For completeness we may wish to write also $G_n^{zz }$ on the same form,
\begin{eqnarray} \label{eq4}
G_n^{zz}  = \frac{1}{4}\left\langle {{\phi _1}{\phi _2}{\phi _3}{\phi_4}}\right\rangle,
\end{eqnarray}
where $\phi_1 = A_1, \phi_2 = B_1, \phi_3 = A_{n+1}, \phi_4 = B_{n+1}$.

The $2n$-point functions [4-point functions] in $\phi$-operators in Eq. (\ref{eq2n}) [Eq. (\ref{eq4})] can be expressed as Pfaffians by the use of the Wick theorem [\onlinecite{Fubini}]. 
For $G_n^{\alpha \beta}$ in Eq. (\ref{eq2n}) one has
\begin{multline} \label{pf2n}
G_n^{\alpha \beta} \\
=  D_n^{\alpha \beta } \mbox{pf} \left( {\begin{array}{*{20}{c}}
{\left\langle {{\phi _1}{\phi _2}} \right\rangle }&{\left\langle {{\phi _1}{\phi _3}} \right\rangle }&{\left\langle {{\phi _1}{\phi _4}} \right\rangle }& \cdots &{\left\langle {{\phi _1}{\phi _{2n}}} \right\rangle }\\
{}&{\left\langle {{\phi _2}{\phi _3}} \right\rangle }&{\left\langle {{\phi _2}{\phi _4}} \right\rangle }& \cdots &{\left\langle {{\phi _2}{\phi _{2n}}} \right\rangle }\\
{}&{}&{\left\langle {{\phi _3}{\phi _4}} \right\rangle }& \cdots &{\left\langle {{\phi _3}{\phi _{2n}}} \right\rangle }\\
{}&{}&{}& \ddots & \vdots \\
{}&{}&{}&{}&{\left\langle {{\phi _{2n - 1}}{\phi _{2n}}} \right\rangle }
\end{array}} \right)
\end{multline}
where we have written the skew-symmetric matrix on standard abbreviated form. $G_n^{zz}$ in Eq. (\ref{eq4}) takes the simpler form,
\begin{equation} \label{pf4}
G_n^{zz} = \frac{1}{4}\mbox{pf}  \left( {\begin{array}{*{20}{c}}
{\left\langle {{\phi _1}{\phi _2}} \right\rangle }&{\left\langle {{\phi _1}{\phi _3}} \right\rangle }&{\left\langle {{\phi _1}{\phi _4}} \right\rangle }\\
{}&{\left\langle {{\phi _2}{\phi _3}} \right\rangle }&{\left\langle {{\phi _2}{\phi _4}} \right\rangle }\\
{}&{}&{\left\langle {{\phi _3}{\phi _4}} \right\rangle }\\
\end{array}} \right)
\end{equation}

The Pfaffians for the diagonal spin correlation functions with $\alpha = \beta = x, y$ reduce to Toeplitz determinants [\onlinecite{Barouch2,Barouch3}], well known from the seminal work on the XY chain by Lieb {\em et al.} [\onlinecite{LSM}] and allowing for an analytical solution when there is no time dependence. Results for the XY-chain two-point functions of spins at different sites {\em and} times, including the off-diagonal cases $\alpha \neq \beta$, have also been obtained in the asymptotic limits of infinite spatial or time separation [\onlinecite{JohnsonMcCoy}]. Here, however, we are interested in the equal-time correlations in Eqs. (A1) and (A2) at finite spatial separation and at nonequilibrium following a quantum quench. For this reason we keep the Pfaffians in Eqs. (\ref{pf2n}) and (\ref{pf4}), insert the appropriate expressions for the four types of two-point functions $\langle {{A_l}{A_m}} \rangle, \langle {{B_l}{B_m}} \rangle, \langle {{A_l}{B_m}} \rangle$ and $\langle {{B_l}{A_m}} \rangle$ that appear as elements in the corresponding skew-symmetric matrices, and then compute the Pfaffians numerically. Note that the result that ensues will be exact, with no approximation invoked. (For a recent analytical approach to spin correlations out of equilibrium in a ``coherent ensemble" of the XY chain in a transverse field, see Ref. [\onlinecite{Granet}].)

Writing out the four types of two-point functions in terms of the JW fermionic operators, we have
\begin{eqnarray}  \label{t4}
\langle {{A_l}{A_m}} \rangle & \!= \!& \langle {a_l^\dag a_m^\dag\rangle \! +\! \langle{a_l}{a_m}\rangle \!+\!\langle a_l^\dag {a_m}\rangle \!+\! \langle{a_l}a_m^\dag }\rangle \\
\langle {{B_l}{B_m}} \rangle &\!= \!& \langle {a_l^\dag a_m^\dag \rangle \!+\! \langle{a_l}{a_m}\rangle \!-\! \langle a_l^\dag {a_m}\rangle \!-\! \langle{a_l}a_m^\dag } \rangle \\
\langle {{A_l}{B_m}} \rangle & \!= \!& \langle {a_l^\dag a_m^\dag\rangle \! - \!\langle{a_l}{a_m}\rangle \!-\! \langle a_l^\dag {a_m}\rangle \!+\! \langle{a_l}a_m^\dag } \rangle \\
\langle {{B_l}{A_m}} \rangle & \!\!= \!\!& \langle {a_l^\dag a_m^\dag\rangle \! +\! \langle a_l a_m\rangle \!-\! \langle a_l^\dag a_m\rangle \!-\! \langle a_l a_m^\dag } \rangle  
\end{eqnarray}

Let us begin by analyzing $\langle a_{\l}^{\dagger} a_{m}\rangle$. Since $1 \!\le\! \l \!\le \!m$ in (A8) - A(11), we can write $\langle a_{\l}^{\dagger} a_{m}\rangle$ as $\langle a_{\l}^{\dagger} a_{\l+r}\rangle$ with $r$ a non-negative integer. As a point of reference we consider the pre-quench expectation value $\langle \Psi_{0;h_1} |a_{\l}^{\dagger} a_{\l+r}  | \Psi_{0;h_1} \rangle$,
with $|\Psi_{0;h_1}\rangle$ the ground state of the XY chain with $h_1$ the magnitude of the transverse field. Taking advantage of translational invariance to write $\langle \Psi_{0;h_1} |a_{\l}^{\dagger} a_{\l+r}  | \Psi_{0;h_1} \rangle = \frac{1}{N}\sum_{\l} \langle \Psi_{0;h_1} |a_{\l}^{\dagger} a_{\l+r}  | \Psi_{0;h_1} \rangle$ and then Fourier transforming, one obtains
\begin{multline} \label{exp1a}
\langle \Psi_{0;h_1} |a_{\l}^{\dagger} a_{\l+r}  | \Psi_{0;h_1} \rangle  \\ = \frac{1}{N}\sum_{k \in BZ}\langle  \Psi_{0;h_1} | a_k^{\dagger}a_k |  \Psi_{0;h_1} \rangle \cos(kr),
\end{multline}
using that the imaginary part of the sum vanishes. 

We now introduce time-dependence into the two-point function by conceiving a quench at time $t=0$ and letting the post-quench Hamiltonian ${\cal H}_2$ with $h \!=\! h_2$ time-evolve the initial state 
$|\Psi_{0;h_1}\rangle\!:\! |\Psi_{0;h_1}(t)\rangle \!=\! e^{-i{\cal H}_2t}|\Psi_{0;h_1}\rangle$; cf. Eq. (\ref{eq2}). Carrying out a Bogoliubov transformation $a_k \!=\! \cos(\theta_k^{(2)})\beta_k + i\sin(\theta_k^{(2)})\beta_k^{\dagger}$ and using our compact notation for time-dependent expectation values (here adapted to the time-evolved ground state $|\Psi_{0;h_1}(t)\rangle$, i.e., $\langle \ldots \rangle := \langle \Psi_{0;h_1}(t) | \ldots | \Psi_{0;h_1}(t) \rangle$), it follows from Eq. (\ref{exp1a}) that   
\begin{eqnarray} \label{exptime}
& & \hspace{-0.25cm}\langle a_{\ell}^{\dagger} a_{\ell +r}\rangle \nonumber \\
&=&\!\!\frac{1}{N}\sum_{k\in BZ} \Big( \langle \beta_k^{\dagger} \beta_k\rangle \cos^2(\theta_k^{(2)}) + \langle \beta_{-k} \beta_{-k}^{\dagger}\rangle \sin^2(\theta_k^{(2)}) \nonumber \\
& & \hspace{1.2cm} + i\big(\langle \beta_k^{\dagger} \beta_{-k}^{\dagger}\rangle e^{2i\varepsilon_k^{(2)}\!t} - \langle \beta_{-k} \beta_k\rangle e^{-2i\varepsilon_k^{(2)}\!t}\big) \nonumber\\
& & \hspace{1.2cm}  \times\cos(\theta_k^{(2)}) \sin(\theta_k^{(2)}) \Big)\cos(kr), \\ \nonumber
\end{eqnarray}
with $\theta_k^{(2)} = -\delta\sin(k)/(\cos(k) + h_2)$ the Bogoliubov angle which diagonalizes ${\cal H}_2$. We have here used that
\begin{equation}
e^{i{\cal H}_2t}\beta_ke^{-i{\cal H}_2t} = \beta_k e^{-i\varepsilon_k^{(2)}\!t}
\end{equation}
with $\varepsilon_k^{(2)} = \sqrt{(\cos(k) + h_2)^2 + \delta^2\sin^2(k)}$, 
exploiting the diagonalized form of the post-quench Hamiltonian, ${\cal H}_2 = \sum_k \varepsilon_k^{(2)}(\beta_k^{\dagger} \beta_k - \frac{1}{2})$. 

Next, we rewrite Eq. (\ref{exptime}) in terms of the Bogoliubov operators $\alpha_k$ and $\alpha_k^{\dagger}$ introduced in Sec. II, related to $\beta_k$ by $\beta_k = \cos(\Phi_k)\alpha_k - i\sin(\Phi_k)\alpha_{-k}^{\dagger}$, with $\Phi_k = \theta_k^{(2)} - \theta_k^{(1)}$. Here $\theta_k^{(1)}$ is the Bogoliubov angle which diagonalizes the pre-quench Hamiltonian ${\cal H}_1$, defined by Eq. (2) with $h=h_1$. Using that the pre-quench ground state $|\Psi_{0;h_1} \rangle$ (assumed to be normalized) serves as a vacuum state for $\alpha_k$, i.e., $\alpha_k|\Psi_{0;h_1}\rangle = 0$, a tedious but straightforward calculation finally returns an expression for $\langle a_{\ell}^{\dagger} a_{\ell +n}\rangle$ with the Bogoliubov quasiparticle two-point functions eliminated: 
\begin{eqnarray} \label{A7}
\langle a_{\ell}^{\dagger} a_{\ell +n}\rangle = -\frac{1}{2N}\sum_{k\in BZ} \Big(\sin(2\Phi_k)\sin(2\theta_k^{(2)})\cos(2\varepsilon_k^{(2)}t)  \nonumber \\
+ \cos(2\Phi_k)\cos(2\theta_k^{(2)})\Big)\!\cos(kn). \nonumber \\ 
\end{eqnarray} 

Repeating the calculational steps above for the other two-point functions $\langle a_l^\dagger a_{m}^\dagger \rangle, \langle{a_l}{a_m}\rangle$, and $\langle {a_l}a_{m}^\dagger \rangle$ that appear in Eqs. (A8)-(A11), putting $m = \l +r$, we obtain the desired closed expressions for the expectation values that enter into the Pfaffians in Eqs. (\ref{pf2n}) and (\ref{pf4}):
\begin{eqnarray}  \label{pfelements}
\langle A_l A_{l+r} \rangle &=& \langle B_l B_{l+r} \rangle \\
&=& \frac{i}{N}\sum\limits_k \!\sin(kr)\sin (2\Phi _k)\sin (2\varepsilon _k^{(2)}t), \nonumber \\
\langle A_l B_{l+r} \rangle  &=& \!\frac{1}{N}\sum \limits_k \!\big(\cos(2\theta _k^{(2)} + kr) \cos(2\Phi _k)  \\
&+& \sin(2\Phi _k) \sin(2\theta _k^{(2)}+ kr) \cos(2\varepsilon _k^{(2)}t) \big), \nonumber \\
\langle B_l A_{l+r} \rangle  &=&  \! -\frac{1}{N}\sum \limits_k \!\big(\cos(2\theta _k^{(2)} - kr) \cos(2\Phi _k)  \\
&+& \sin(2\Phi _k) \sin(2\theta _k^{(2)}- kr) \cos(2\varepsilon _k^{(2)}t) \big), \nonumber 
\end{eqnarray}
where $\l \!= \!1,2,..., n;~r=0,1,2,...,n \ [\l \!=\! 1, n\!+\!1; r\!=\!0,n]$ in (\ref{pf2n}) [(\ref{pf4})]. 
Inserting (A16) - (A18) into (\ref{pf2n}) and (\ref{pf4}), the Pfaffians can now be computed, yielding exact results for the post-quench spin correlations in Eqs. (A1) and (A2). For this purpose we have used a Python code which implements an algorithm in Ref. [\onlinecite{Wimmer}]. Note that  in (A16) we have used that for both  $\langle A_l A_{l+r} \rangle $ and $\langle B_l B_{l+r} \rangle$ there is no situation in which  $r=0$, otherwise, $\langle A_l A_l \rangle =-\langle B_l B_l \rangle$=1.}
\subsection*{}


\begin{thebibliography}{99}

\bibitem{Radcliffe}
J. M. Radcliffe, J. Phys. A: Gen. Phys. {\bf 4}, 313 (1971).

\bibitem{Kitagawa}
M. Kitagawa, and M. Ueda, Phys. Rev. A {\bf 47}, 5138 (1993).

\bibitem{Wineland}
D. J. Wineland, J. J. Bollinger, W. M. Itano, F. L. Moore, D. J. Heinzen, Phys. Rev. A {\bf 46}, R6797 (1992); 
D. J. Wineland, J. J. Bollinger, W. M. Itano,  and D. J. Heinzen, Phys. Rev. A {\bf 50}, 67 (1994).

\bibitem{Ma1}
For a review, see J. Ma, X. Wang, C. P. Sun, and F. Nori, Phys. Rep. {\bf 509}, 89 (2011). 

\bibitem{Sorensen}
A. S$\o$rensen, L. Duan, J. Cirac, and  P. Zoller, Nature {\bf 409}, 63 (2001).

\bibitem{Wang}
X. Wang,  and B. C. Sanders, Phys. Rev. A {\bf 68}, 012101 (2003).

\bibitem{Toth}
G. Toth,  Ch. Knapp, O. Guhne,  and H. J. Briegel, Phys. Rev. Lett. {\bf 99}, 250405 (2007); 
G. Toth,  Ch. Knapp, O. Guhne,  and H. J. Briegel, Phys. Rev. A {\bf 79}, 042334 (2009).

\bibitem{Guehne}
O. Guehne,  and G. Toth, Phys. Rep. {\bf 474}, 1 (2009).

\bibitem{Giovannetti}
V. Giovannetti, S. LIoyd,  and L. Maccone, Phys. Rev. Lett. {\bf 96}, 010401 (2006);  Nat. Photon.  {\bf 5}, 222 (2011).

\bibitem{Hald}
J. Hald, J. L. S$\o$rensen, C. Schori, and E. S. Polzik, Phys. Rev. Lett. {\bf 83}, 1319 (1999); 
T. Fernholz, H. Krauter, K. Jensen, J. F. Sherson, A. S. S$\o$rensen, and E. S. Polzik, Phys. Rev. Lett. {\bf 101}, 073601 (2008).

\bibitem{Orzel}
 C. Orzel, A. Tuchman, M. Fenselau, M. Yasuda, and M. Kasevich, Science {\bf 291}, 2386 (2001).
 
\bibitem{Meyer}
V. Meyer, M. A. Rowe, D. Kielpinski, C. A. Sackett, W. M. Itano, C. Monroe, and D. J. Wineland, Phys. Rev. Lett. {\bf 86}, 5870 (2001). 

\bibitem{Smith}
G. A. Smith, S. Chaudhury, A. Silberfarb, I. H. Deutsch, and P. S. Jessen, Phys. Rev. Lett. {\bf 93}, 163602 (2004). 

\bibitem{Takano}
T. Takano, M. Fuyama, R. Namiki, and Y. Takahashi, Phys. Rev. Lett. {\bf 102}, 033601 (2009).

\bibitem{Gross1}
C. Gross, T. Zibold, E. Nicklas, J. Esteve,  and M. K. Oberthaler, Nature {\bf 464}, 1165 (2010).

\bibitem{Riedel}
M. F. Riedel, P. Bohi, Y. Li, T. W. Hansch, A. Sinatra,  and P. Treutlein, Nature {\bf 464}, 1170 (2010).

\bibitem{Hamley}
C. D. Hamley, C. S. Gerving, T. M. Hoang, E. M. Bookjans,  and M. S. Chapman, Nat. Phys. {\bf 8}, 305 (2012).

\bibitem{Auccaise}
R. Auccaise, A. G. Araujo-Ferreira, R. S. Sarthour, I. S. Oliveira, T. J. Bonagamba,  and I. Roditi, Phys. Rev. Lett. {\bf 114}, 043604 (2015).

\bibitem{Hosten}
O. Hosten, N. J. Engelsen, R. Krishnakumar,  and M. A. Kasevich, Nature {\bf 529}, 505 (2016).

\bibitem{Orioli}
A. P. Orioli, A. Signoles, H. Wildhagen, G. Gunter, J. Berges, S. Whitlock,  and M. Weidemuller, Phys. Rev. Lett. {\bf 120}, 063601 (2018).

\bibitem{Pezze}
L. Pezz\`e, A. Smerzi, M. K. Oberthaler, R. Schmied, and P. Treutlein, Rev. Mod. Phys. {\bf 90}, 035005 (2018).

\bibitem{Braverman}
B. Braverman, A. Kawasaki, E. Pedrozo-Penafiel, S. Colombo, C. Shu, Z. Li, E. Mendez, M. Yamoah, L. Salvi, D. Akamatsu, Y. Xiao,  and V. Vuletic,  
Phys. Rev. Lett. {\bf 122}, 223203 (2019).

\bibitem{Bao}
H. Bao, J. Duan, S. Jin, X. Lu, P. Li, W. Du, M. Wang, I. Novikova, E. E. Mikhailov, K.-F. Zhao, K. M$\o$lmer, H. Shen,  and Y. Xiao, Nature {\bf 581}, 159 (2020).

\bibitem{Sorensen1}
A. S$\o$renson,  and K. M$\o$lmer, Phys. Rev. Lett. {\bf 83}, 2274 (1999).

\bibitem{Duan}
L.-M. Duan, A. S. S$\o$rensen,  J. I. Cirac,  and  P. Zoller,  Phys. Rev. Lett. {\bf 85}, 3991 (2000);

\bibitem{SorensenNature}
A. S. S$\o$rensen, L. -M. Duan, J. I. Cirac, and  P. Zoller,  Nature (London) {\bf 409}, 63 (2001).

\bibitem{Law}
C. K. Law, H. T. Ng,  and P. T. Leung, Phys. Rev. A {\bf 63}, 055601 (2001).

\bibitem{Wang2}
X. Wang, A. S. S$\o$rensen,  and K. M$\o$lmer, Phys. Rev. A {\bf 64}, 053815 (2001).

\bibitem{Sorensen2}
A. S. S$\o$rensen,  and K. M$\o$lmer, Phys. Rev. Lett. {\bf 86}, 4431 (2001).

\bibitem{Wang1}
X. Wang,  and B. C. Sanders, Phys. Rev. A {\bf 68}, 012101 (2003).

\bibitem{Ma2}
J. Ma,  and X. Wang, Phys. Rev. A {\bf 80}, 012318 (2009).

\bibitem{Wang3}
X. Wang, A. Miranowicz, Y.-X. Liu, C. P. Sun,  and F. Nori, Phys. Rev. A {\bf 81}, 022106 (2010).

\bibitem{Liu}
W.-F. Liu, J. Ma,  and X. Wang, J. Phys. A: Math. Theor. {\bf 46}, 045302 (2013).   

\bibitem{Abad}
T. Abad, K. M$\o$lmer,  and V. Karimipour, Phys. Rev. A {\bf 96}, 042337 (2017).

\bibitem{Xu1}
P. Xu, H. Sun, S. Yi,  and W. Zhang, Sci. Rep. {\bf 7}, 14102 (2017).

\bibitem{Frerot}
I. Frerot,  and T. Roscilde, Phys. Rev. Lett. {\bf 121}, 020402 (2018).

\bibitem{Balazadeh}
L. Balazadeh, G. Najarbashi,  and A. Tavana, Sci. Rep. {\bf 8}, 17789 (2018).

\bibitem{Kaubruegger}
R. Kaubruegger, P. Silvi, C. Kokail, R. van Bijnen, A. M. Rey, J. Ye, A. M. Kaufman, and  P. Zoller, Phys. Rev. Lett. {\bf 123}, 260505 (2019).

\bibitem{Schulte}
M. Schulte, C. Lisdat, P. O. Schmidt, U. Sterr,  and K. Hammerer, Nat. Commun. {\bf 11}, 5955 (2020).

\bibitem{Qin}
{\color{black} W. Qin, Y.-H. Chen, X. Wang, A. Miranowicz, and F. Nori, Nanophotonics {\bf 9}, 4853 (2020).}

\bibitem{Comparin2}
{\color{black} T. Roscilde, F. Mezzacapo, and T. Comparin,
Phys. Rev. A {\bf 104}, L040601 (2021).}
 
\bibitem{Zibold}
T. Zibold, E. Nicklas, C. Gross, and M. K. Oberthaler, Phys. Rev. Lett. {\bf 105}, 204101 (2010).

\bibitem{Martin}
M. J. Martin, M. Bishof, M. D. Swallows, X. Zhang,
C. Benko, J. von-Stecher, A. V. Gorshkov, A. M. Rey,
and J. Ye, Science {\bf 341}, 632 (2013).

\bibitem{Britton}
J. W. Britton, B. C. Sawyer, A. C. Keith, C.-C. J.
Wang, J. K. Freericks, H. Uys, M. J. Biercuk, and J. J.
Bollinger, Nature {\bf 484}, 489 (2012).

\bibitem{Bohnet}
J. G. Bohnet, B. C. Sawyer, J. W. Britton, M. L. Wall,
A. M. Rey, M. Foss-Feig, and J. J. Bollinger, Science {\bf 352}, 1297 (2016).

\bibitem{Baumann}
K. Baumann, C. Guerlin, F. Brennecke, and T. Esslinger, Nature {\bf 464}, 1301 (2010).

\bibitem{Ritsch}
H. Ritsch, P. Domokos, F. Brennecke, and T. Esslinger,
Rev. Mod. Phys. {\bf 85}, 553 (2013).

\bibitem{Davis}
E. J. Davis, G. Bentsen, L. Homeier, T. Li, and
M. H. Schleier-Smith, Phys. Rev. Lett. {\bf 122}, 010405 (2019).

\bibitem{Hazzard}
K. R. A. Hazzard, M. van den Worm, M. Foss-Feig, S. R. Manmana, E. G. Dalla Torre, T. Pfau, M. Kastner,  and A. M. Rey, Phys. Rev. A {\bf 90}, 063622 (2014).

\bibitem{Feig}
{\color{black} M. Foss-Feig, Z.-X. Gong, A. V. Gorshkov, and C. W. Clark, arXiv:1612.07805 [cond-matt.quant-gas].}

\bibitem{Lerose}
A. Lerose,  and S. Pappalardi,  Phys. Rev. Research {\bf 2}, 012041(R) (2020).

\bibitem{Gietka}
{\color{black} K. Gietka, A. Usui, J. Deng, and T. Busch, Phys. Rev. Lett. {\bf 126}, 160402 (2021).}  

\bibitem{Perlin}
M. A. Perlin, C. Qu, and A. M. Rey, Phys. Rev. Lett. {\bf 125}, 223401 (2020). 

\bibitem{Comparin1}
{\color{black} T. Comparin, F. Mezzacapo, and T. Roscilde, arXiv:2103.07354.} 

{\color{black}
\bibitem{Cazalilla}
M. A. Cazalilla and A. M. Rey, Rep. Prog. Phys. {\bf 77}, 124401 (2014).

\bibitem{Gross2}
C. Gross and I. Bloch, Science {\bf 357}, 995 (2017).

\bibitem{Browaeys}
A. Browaeys and T. Lahaye, Nat. Phys. {\bf 16}, 132 (2020).}

\bibitem{Eisert}
For reviews, see J. Eisert, M. Friesdorf, and C. Gogolin, Nat. Phys. {\bf 11}, 124 (2015); 
T. Langen, R. Geiger, and J. Schmiedmayer, Annu. Rev. Condens. Matter Phys. {\bf 6}, 201 (2015).

\bibitem{Barouch1}
E. Barouch, B. M. McCoy, and M. Dresden, Phys. Rev. A {\bf 2}, 1075 (1970).

\bibitem{Igloi}
{\color{black} B. Blass, H. Rieger, and F. Igl\'{o}i, Europhys. Lett. {\bf 99}, 30004 (2012).} 

\bibitem{Essler}
For a review, see F. H. L. Essler and M. Fagotti, J. Stat. Mech. 064002 (2016).

\bibitem{Katsura}
S. Katsura, Phys. Rev. {\bf 127}, 1508 (1962).

\bibitem{Niemeijer}
T. Niemeijer, Physica {\bf 36}, 377(1967); T. Niemeijer, Physica {\bf 39}, 313 (1968).

\bibitem{Xu2}
K. Xu, Z.-H. Sun, W. Liu, Y.-R. Zhang, H. Li, H. Dong, W. Ren, P. Zhang, F. Nori, D. Zheng, H. Fan,  and H. Wang,  Sci. Adv. {\bf 6}, 4935 (2020).

\bibitem{Hamma}
J. H\"app\"ol\"a, G. B. Hal\'asz and A. Hamma, Phys. Rev. A {\bf 85}, 032114 (2012).

\bibitem{Henrik}
R. Jafari and H. Johannesson, Phys. Rev. Lett.  {\bf 118}, 015701 (2017).

\bibitem{Delgado}
{\color{black} R. Jafari, H. Johannesson, A. Langari, and M. A. Martin-Delgado, Phys. Rev. B {\bf 99}, 054302 (2019).}

\bibitem{Mishra}
{\color{black} U. Mishra, H. Cheraghi, S. Mahdavifar, R. Jafari, and A. Akbari,
Phys. Rev. A {\bf 98}, 052338 (2018).}

\bibitem{Hadi2}
H. Cheraghi  and S. Mahdavifar, Phys. Rev. B  {\bf 102}, 024304 (2020).

\bibitem{Akbari}
{\color{black} R. Jafari and A. Akbari,
Phys. Rev. A {\bf 101}, 062105 (2020).}

\bibitem{Kurmann}
J. Kurmann, H. Thomas, and G. M\"uller, Physica A {\bf 112}, 235 (1982); 
G. M\"uller and R. E. Shrock, Phys. Rev. B {\bf 32}, 5845 (1985).

\bibitem{Adesso}
S. M. Giampaolo, G. Adesso, and F. Illuminati, Phys. Rev. Lett. {\bf 100}, 197201 (2008).

\bibitem{LSM}
E. Lieb, T. Schultz, and D. Mattis, Ann. Phys. {\bf 16}, 407 (1961).

\bibitem{FranchiniBook}
F. Franchini, {\em An Introduction to Integrable Techniques for One-Dimensional Quantum Systems} (Springer, 2017).

\bibitem{Schreiber}
M. Schreiber, S. S. Hodgman, P. Bordia, H. P. L\"uschen, M. H. Fischer, R. Vosk, E. Altman, U. Schneider, and  I. Bloch, Science {\bf 349}, 842 (2015).

\bibitem{Choi}
S. Choi, J. Choi, R. Landig, G. Kucsko, H. Zhou, J. Isoya, F. Jelezko, S. Onoda, H. Sumiya, V. Khemani, C. von Keyserlingk, N. Y. Yao, E. Demler and M. D. Lukin, Nature {\bf 543}, 221 (2017).

\bibitem{Gorg}
F. G\"org, K. Sandholzer, J. Minguzzi, R. Desbuquois, M. Messer and T. Esslinger, Nat. Phys. {\bf 15}, 1161-1167 (2019).

\bibitem{Mitra}
For a review, see A. Mitra, Annu. Rev. Condens. Matter Phys. {\bf 9}, 245 (2018).

\bibitem{Barouch2}
E. Barouch and B. M. McCoy, Phys. Rev. A {\bf 3}, 786 (1971). 

\bibitem{Fubini}
{\color{black} E. R. Caianello and S. Fubini, Nuovo Cim. {\bf 9}, 1218 (1952).} 

\bibitem{HadiNumerical}
H. Cheraghi, {\em unpublished}.

\bibitem{Osterloh}
A. Osterloh, L. Amico, G. Falci, and R. Fazio, Nature {\bf 416}, 608 (2002).

\bibitem{Osborne}
T. J. Osborne and M. A. Nielsen, Phys. Rev. A {\bf 66}, 032110 (2002).

\bibitem{Vidal}
G. Vidal, J. I. Latorre, E. Rico, and A. Kitaev, Phys. Rev. Lett. {\bf 90}, 227902 (2003). 

\bibitem{Knap}
B. \v{Z}unkovi\v{c}, M. Heyl, M. Knap, and A. Silva, Phys. Rev. Lett. {\bf 120}, 130601 (2018).

\bibitem{Zhou}
B. Zhou, C. Yang, and S. Chen, Phys. Rev. B  {\bf 100}, 184313 (2019).

\bibitem{Hadi3}
H. Cheraghi and S. Mahdavifar, Sci. Rep. {\bf 10}, 4407 (2020).  

\bibitem{Gorin}
T. Gorin, T. Prosen, T. H. Seligman, and M. Znidaric, Phys. Rep. {\bf 435}, 33 (2006).

\bibitem{Quan}
H. T. Quan, Z. Song, X. F. Liu, P. Zanardi, and C. P. Sun, Phys. Rev. Lett. {\bf 96}, 140604 (2006).

\bibitem{Yuan}
Z.-G. Yuan, P. Zhang, and S.-S. Li, Phys. Rev. A {\bf 75}, 012102 (2007).

\bibitem{Rossini}
D. Rossini, T. Calarco, V. Giovannetti, S. Montangero, and R. Fazio, Phys. Rev. A {\bf 75}, 032333 (2007).

\bibitem{Zhong}
M. Zhong and P. Tong, Phys. Rev. A {\bf 84}, 052105 (2011).

\bibitem{Fago}
M. Fagotti and P. Calabrese, Phys. Rev. A 78, 010306(R) (2008).

\bibitem{Dubail}
J. M. St\'ephan and J. Dubail, J. Stat. Mech.: Theory Exp. (2011) P08019.

\bibitem{Najafi}
K. Najafi and M. A. Rajabpour, Phys. Rev. B  {\bf 96}, 014305 (2017). 

\bibitem{Brune}
{\color{black} T. L. Nguyen, J.-M. Raimond, C. Sayrin, R. Cortinas, T. Cantat-Moltrecht, F. Assemat, I. Dotsenko, S. Gleyzes, S. Haroche, G. Roux, T. Jolicoeur, and M. Brune,
Phys. Rev. X {\bf 8}, 011032 (2018).}

\bibitem{Mallesh}
K. S. Mallesh, S. Sirsi, M. A. A. Sbaih, P. N. Deepak, and G. Ramachandran, J. Phys. A : Math. Gen. {\bf 34}, 3293 (2001).

\bibitem{Barouch3}
{\color{black} E. Barouch and B. M. McCoy, Phys. Rev. A {\bf 3}, 2137 (1971).} 

\bibitem{JohnsonMcCoy}
{\color{black} J. D. Johnson and B. M. McCoy, Phys. Rev. A {\bf 4}, 2314 (1971).} 

\bibitem{Granet}
{\color{black} E. Granet, H. Dreyer, and F. H. L. Essler, arXiv:2106.08359.} 

\bibitem{Wimmer}
{\color{black} M. Wimmer, ACM Trans. Math. Softw. {\bf 38}, 30 (2012).}



\end{thebibliography}
\end{document}